\documentclass[sigconf]{acmart}

\usepackage[english]{babel}
\usepackage[ruled,vlined]{algorithm2e}
\usepackage{enumitem}
\setlist{nolistsep}
\usepackage{gensymb}
\graphicspath{{gfx/}}

\newcommand{\fref}[1]{Fig.~\ref{#1}}

\newcommand{\ket}[1]{|#1\rangle}

\copyrightyear{2020}
\acmYear{2020}
\setcopyright{rightsretained}
\acmConference[CoNEXT '20]{The 16th International Conference on emerging Networking EXperiments and Technologies}{December 1--4, 2020}{Barcelona, Spain}
\acmBooktitle{The 16th International Conference on emerging Networking EXperiments and Technologies (CoNEXT '20), December 1--4, 2020, Barcelona, Spain}
\acmDOI{10.1145/3386367.3431293}
\acmISBN{978-1-4503-7948-9/20/12}

\begin{document}

\title{Designing a Quantum Network Protocol}

\author{Wojciech Kozlowski, Axel Dahlberg, and Stephanie Wehner}
\affiliation{%
  \institution{QuTech, Delft University of Technology}
  \institution{Kavli Institute of Nanoscience, Delft University of Technology}
  \streetaddress{Lorentzweg 1}
  \postcode{2628 CJ}
  \city{Delft}
  \country{The Netherlands}
}
\email{{w.kozlowski, s.d.c.wehner}@tudelft.nl}

\begin{abstract}
  The second quantum revolution brings with it the promise of a quantum
  internet. As the first quantum network hardware prototypes near completion
  new challenges emerge. A functional network is more than just the physical
  hardware, yet work on scalable quantum network systems is in its infancy. In
  this paper we present a quantum network protocol designed to enable
  end-to-end quantum communication in the face of the new fundamental and
  technical challenges brought by quantum mechanics. We develop a quantum data
  plane protocol that enables end-to-end quantum communication and can serve as
  a building block for more complex services. One of the key challenges in
  near-term quantum technology is decoherence --- the gradual decay of quantum
  information --- which imposes extremely stringent limits on storage times.
  Our protocol is designed to be efficient in the face of short quantum memory
  lifetimes. We demonstrate this using a simulator for quantum networks and
  show that the protocol is able to deliver its service even in the face of
  significant losses due to decoherence. Finally, we conclude by showing that
  the protocol remains functional on the extremely resource limited hardware
  that is being developed today underlining the timeliness of this work.
\end{abstract}



\maketitle

\section{Introduction}

\begin{figure}[t!]
  \centering
  \includegraphics[width=\linewidth]{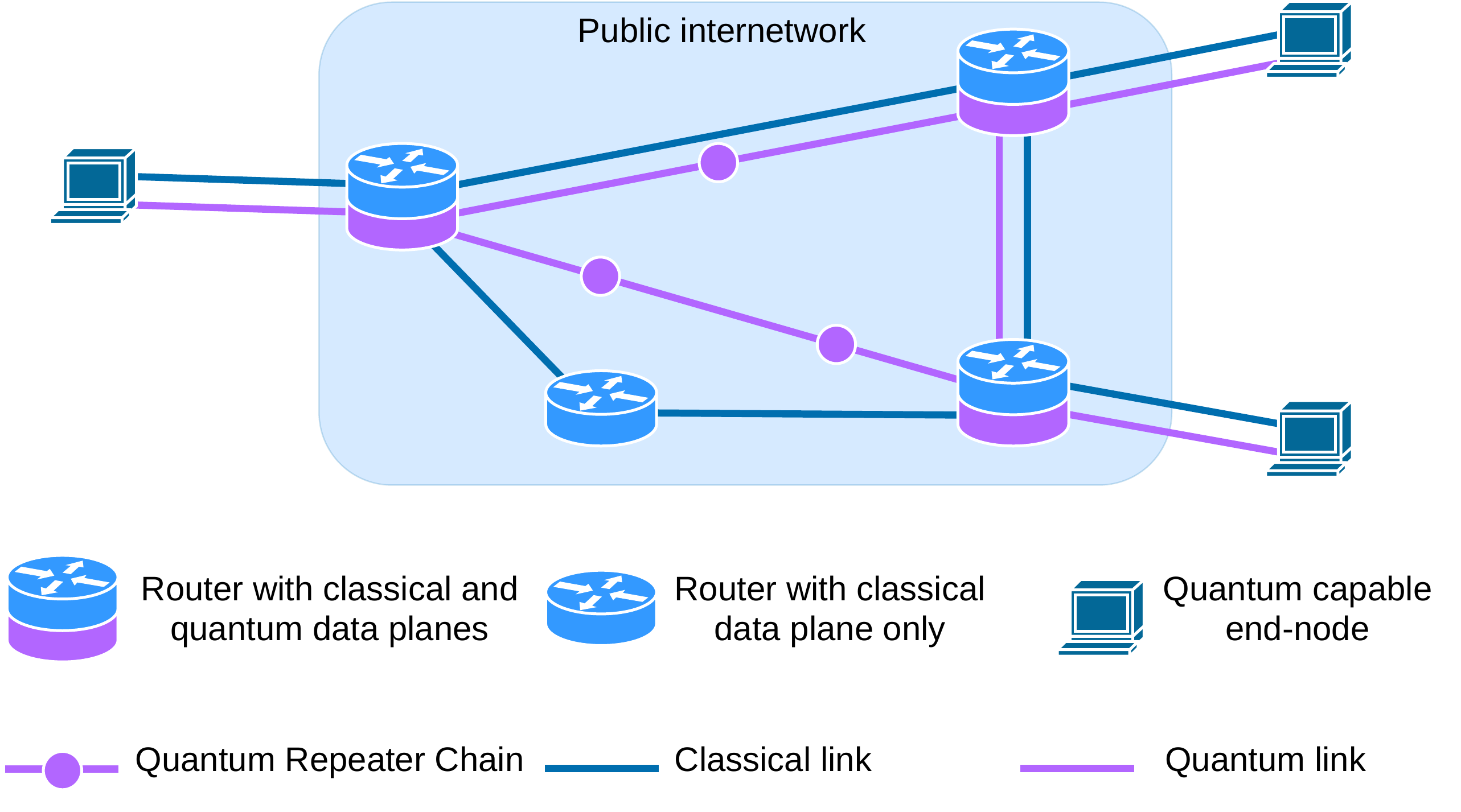}
  \caption{Quantum networks will use existing network infrastructure to
    exchange classical messages for the purposes of running quantum protocols
    as well as the control and management of the network itself. Long-distance
    links will be built using chains of automated quantum repeaters.
  }\label{fig:qnetwork}
\end{figure}

The second quantum revolution is currently unfolding across the scientific
world~\cite{dowling2003quantum}. It brings with it the promise of a quantum
internet, a global network capable of transmitting quantum
data~\cite{wehner2018quantum, kozlowski2019towards}. Quantum networks will
enhance non-quantum (classical) networks (\fref{fig:qnetwork}) and they will
execute protocols that are provably impossible to do classically or that are
more efficient than what is possible classically. This new paradigm enables new
possibilities such as quantum secure communications~\cite{bennett1984quantum,
  ekert1991quantum}, distributed quantum computation~\cite{crepeau2002secure},
secure quantum computing in the cloud~\cite{broadbent2009universal,
  fitzsimons2017unconditionally}, clock
synchronisation~\cite{komar2014quantum}, and quantum-enhanced measurement
networks~\cite{giovannetti2004quantum, gottesman2012longer}. This technology is
developing rapidly with the first inter-city network planned to go online in
the next few years~\cite{qia}.

Quantum communication has been actively researched for many years. Its most
well-known application, quantum key distribution (QKD) is a protocol used for
secure communications~\cite{bennett1984quantum, ekert1991quantum}.
Short-distance QKD networks are already being deployed and studied in
metropolitan environments (e.g.~\cite{sasaki2011field, peev2009secoqc,
  stucki2011long, wang2014field}) and are even commercially available
(e.g.~\cite{giovannetti2004quantum, inagaki2013entanglement,
  diamanti2016practical, fibresystems}). Longer distance QKD networks are
currently possible provided all intermediate nodes are trusted and physically
secure~\cite{salvail2010security, scarani2009security, courtland2016china,
  sasaki2011field}. However, whilst these nodes are capable of exchanging
quantum bits (qubits) with their neighbours, they are not capable of forwarding
them (including by means of entanglement swapping, a method explained later in
this paper). As a result such networks are unable to transmit qubits end-to-end
and thus do not offer end-to-end security.

The next step is to enable long-distance end-to-end communication of qubits.
There are three key challenges in realising this objective: transmission
losses, decoherence, and the no-cloning theorem. Decoherence is the loss of
quantum information due to interactions with the environment and it limits the
lifetime of quantum memories. Typical memory lifetimes in quantum networking
hardware range from a few microseconds to just over a
second~\cite{abobeih2018one} though lifetimes of up to a minute have been
observed in devices disconnected from a network~\cite{bradley201910}. The
no-cloning theorem states that arbitrary quantum data cannot be copied.
Therefore, it is impossible to use standard techniques of amplification or
retransmission to compensate for transmission or decoherence losses. Quantum
error-correcting techniques for quantum repeaters exist~\cite{munro2012quantum,
  fowler2010surface, muralidharan2014ultrafast} which eventually would be able
to compensate for both types of losses~\cite{muralidharan2016optimal}, but they
are extremely demanding in terms of resources and will likely not be feasible
for a few more decades.

An alternative to directly transmitting qubits relies on distributing entangled
pair states. Quantum entanglement is a special state of two or more qubits in
which the individual qubits cannot be described independently of the others, in
principle, across arbitrary distances. It is the key ingredient for
long-distance communication, because one can use an entangled pair of qubits to
teleport an arbitrary data qubit. This bypasses the problem of losses and the
no-cloning theorem, because the entangled pairs can easily be regenerated when
lost as they need only be delivered from a small set of particular states
called the Bell states. Furthermore, this method overcomes transmission losses
as long-range entanglement can be created by ``stitching'' shorter-range pairs
together through a process called entanglement
swapping~\cite{briegel1998quantum} which means that it is not necessary to
transmit qubits directly along the entire path. Entanglement generation between
two directly connected nodes with a quantum memory has been demonstrated at
distances of up to 1.3\,km~\cite{hensen2015loophole} and work is underway to
build a three-node setup and extend the inter-node distances to several
kilometres~\cite{dreau2018quantum, tchebotareva2019entanglement}.

\begin{figure}[t!]
  \centering
  \includegraphics[width=0.9\linewidth]{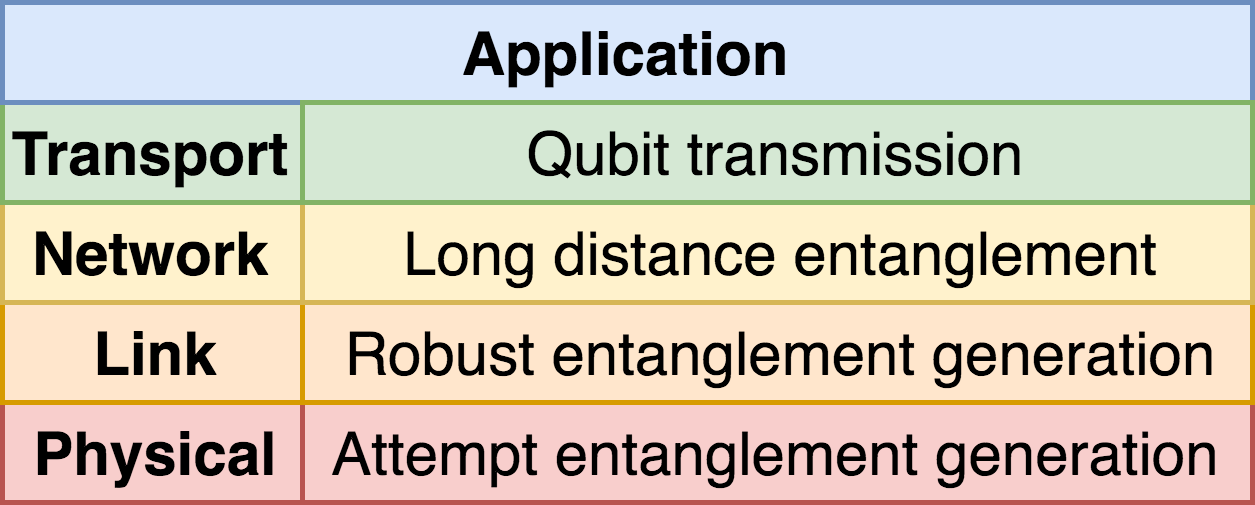}
  \caption{Functional allocation in a quantum network stack from
    Ref.~\cite{dahlberg2019link}. The structure is inspired by the TCP/IP
    stack.}\label{fig:stack}
\end{figure}

In this paper we design a quantum network protocol capable of generating
end-to-end entanglement marking the next step in the development of
long-distance quantum communication networks. The starting point for our work
is a recently proposed protocol for generating link-level
entanglement~\cite{dahlberg2019link}. Going from link-level entanglement to
end-to-end entanglement is a significant leap in complexity as it requires many
new mechanisms that do not exist at the link level. In our protocol we develop
solutions for: (i) coordinating entanglement swapping between multiple nodes in
order to ``stitch'' link-level entanglement into long-range end-to-end
entanglement, (ii) reducing the amount of decoherence experienced by qubits
stored in quantum memory, (iii) compensating for qubits lost due to
decoherence, (iv) ensuring that the final entangled pair is of sufficient
quality to be useful in an application, and many other problems. The result of
our work is a quantum data plane protocol capable of creating end-to-end
entanglement thus enabling long-distance quantum networks. In particular, our
design focuses on ensuring efficient entanglement generation in the face of
short memory lifetimes. At the same time we ensure scalability by designing the
protocol to be a building block for more complex quantum network services
rather than a complete all-in-one solution. Our key research contributions are:

\begin{enumerate}
  \item We design a protocol for generating end-to-end entangled pairs in the
    face of decoherence that fulfils the role of a quantum network layer.
  \item We outline the architecture for the construction of quantum network
    services and design our protocol to fulfil the role of the building block
    in this scheme.
  \item We evaluate the effectiveness of the proposed protocol against
    decoherence in a quantum network simulator.
  \item We show that it remains functional on extremely limited near-term
    hardware justifying its timeliness.
\end{enumerate}

\section{Background and Motivation}\label{sec:background}

\begin{figure}[t!]
  \centering \includegraphics[width=0.9\linewidth]{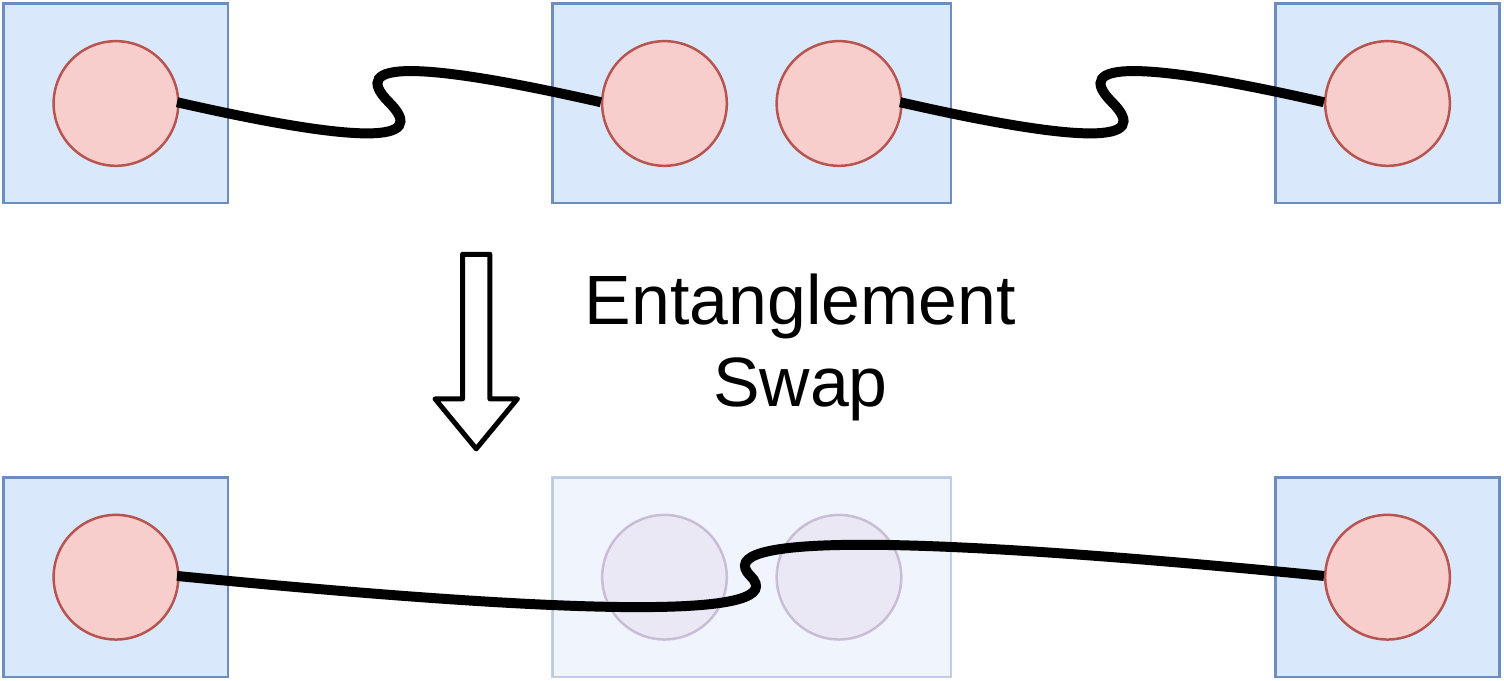}
  \caption{Quantum repeaters create long-distance entanglement by connecting
    short-distance entangled pairs. Initially two entangled pairs each have one
    qubit in the memory of the middle repeater. An entanglement swap is
    performed on these qubits which destroys the entanglement of the two pairs,
    but as a result the remote qubits become entangled.}\label{fig:swap}
\end{figure}

Here, we provide the motivation and justify the timeliness of a quantum network
protocol. We only provide an introduction to the quantum mechanical concepts
necessary to understand the protocol design. Nevertheless, quantum networks are
not new in literature and good introductions to the subject can be found in
Refs.~\cite{van2014quantum, wehner2018quantum, van2013designing,
  dahlberg2019link, kozlowski2019towards}.

\subsection{Motivation}

So far, the generation of long-lived entanglement has been the domain of highly
sophisticated physics experiments. However, real deployments of quantum
networks are around the corner with the first inter-city network scheduled to
go online within the next few years~\cite{qia}. Much essential work is being
done to build quantum hardware to make this
possible~\cite{sangouard2011quantum, munro2015inside, wehner2018quantum,
  bradley201910, tchebotareva2019entanglement} and we are now entering a new
phase of development where we need to learn how to build quantum communication
\emph{systems}. Work in this field has been slowly emerging over the last few
years (see e.g.~\cite{van2008system, santra2019quantum, khatri2019practical,
  munro2015inside, matsuo2019quantum, caleffi2020quantum}). Recently, a
proposal for a quantum network stack inspired by TCP/IP has been put forward
(\fref{fig:stack}) along with a link layer protocol that provides a robust
entanglement generation service between directly connected
nodes~\cite{dahlberg2019link}. Here, we go one level up this network stack and
achieve the next step in quantum connectivity, a quantum network layer protocol
capable of providing long distance end-to-end entanglement between any pair of
nodes in the network.

\subsection{Entanglement Swapping}\label{sec:swap}

In light of the the no-cloning theorem, decoherence, and transmission losses
how can entangled qubits be practically distributed if we cannot use
amplification or retransmissions? In 1998 Briegel et
al.~\cite{briegel1998quantum} proposed a solution whereby \emph{quantum
repeaters} create long-distance entanglement by connecting a string of
short-distance entangled pairs of qubits through a process called entanglement
swapping, shown in \fref{fig:swap}. Therefore, a practical scheme for
distributing entanglement may combine a scheme for generating short-distance
entangled pairs, such as a quantum link layer protocol~\cite{dahlberg2019link}
which wraps the physical mechanism~\cite{humphreys2018deterministic,
  inlek2017multispecies, reiserer2015cavity} for pair generation, with
entanglement swapping at quantum repeaters.

Despite the quantum nature of the underlying physical processes, quantum
networks will require classical connectivity between all the quantum nodes as
shown in \fref{fig:qnetwork} for the exchange of control messages. Most
notably, entanglement swapping as shown in \fref{fig:swap} requires the middle
node to send a message to at least one of the other nodes for the entanglement
to be useful\footnote{The entanglement swap results in one of four possible
entangled states, but which state is produced is fundamentally random. The node
that performed the entanglement swap will obtain two bits of information
indicating which state was produced. Without this information the remote nodes
do not know what state they share rendering it useless to any application.}.
Furthermore, just like classical networks, quantum networks will need control
and management protocols which will also use the classical channels.

\subsection{Fidelity and Decoherence}

Next to standard measures like throughput and latency, a key parameter in a
quantum network is a quantity called \emph{fidelity}~\cite{dahlberg2019link}.
Fidelity is a purely quantum metric with no classical equivalent. Its value
lies between 0 and 1 and it quantifies the quality of the state in terms of how
``close'' it is to the desired state (a fidelity of 1 means it is in the
desired state, a value below 0.5 means that the state is no longer usable). It
is important to note that unlike in classical networks where data must be
delivered error-free, quantum applications are able to operate with imperfect
quantum states --- as long as the fidelity is above an application-specific
threshold (for basic QKD the threshold fidelity is about $0.8$).

Decoherence is the gradual degradation of qubit quality over time and will
cause the fidelity to decrease. Decoherence is one of the key challenges in
quantum networks as it puts extremely stringent limits on how long qubits can
be held in memory before they need to be used. In current experimental
hardware, these times are of the order of few
milliseconds~\cite{dahlberg2019link, humphreys2018deterministic}, but memories
in similar devices disconnected from a network have shown lifetimes of up to
one minute~\cite{bradley201910}.

Quantum state fidelity in a network is lost in several ways:

\begin{enumerate}[label=(P\arabic*)]
  \item Short-range pairs generated on a link are imperfect.
  \item Swapping imperfect pairs results in a pair of lower fidelity even if
    the physical operations are noiseless.
  \item Imperfect implementations of quantum gates reduce fidelity whenever any
    qubit is processed.
  \item Decoherence degrades a quantum state's fidelity while the qubits are
    stored in memory.
\end{enumerate}

Whilst the fidelity of a short-distance pair generated on a link (P1) is
ultimately the property of the hardware, some implementations are able to vary
the fidelity of the produced pairs though higher fidelities come at the cost of
reduced rates~\cite{dahlberg2019link}. The issue in (P2) is a fundamental
property of entanglement swapping and the only way to ensure that the output
state is sufficiently good is to feed sufficiently high quality states into the
swap. (P3) is similar, but can also be addressed by improving the hardware
which is out of scope for a network protocol. Finally, decoherence (P4) can be
addressed at the protocol level by minimising the time qubits spend idling in
memory. Therefore, in our design we focus on addressing two key questions: (i)
how does the protocol know what fidelity to request on the individual links to
ensure a sufficiently high end-to-end fidelity after all the operations
complete, and more importantly (ii) how to minimise decoherence by reducing the
amount of time qubits sit idly in memory.

\subsection{Quantum Node Architecture}

\begin{figure}[t!]
  \centering \includegraphics[width=\linewidth]{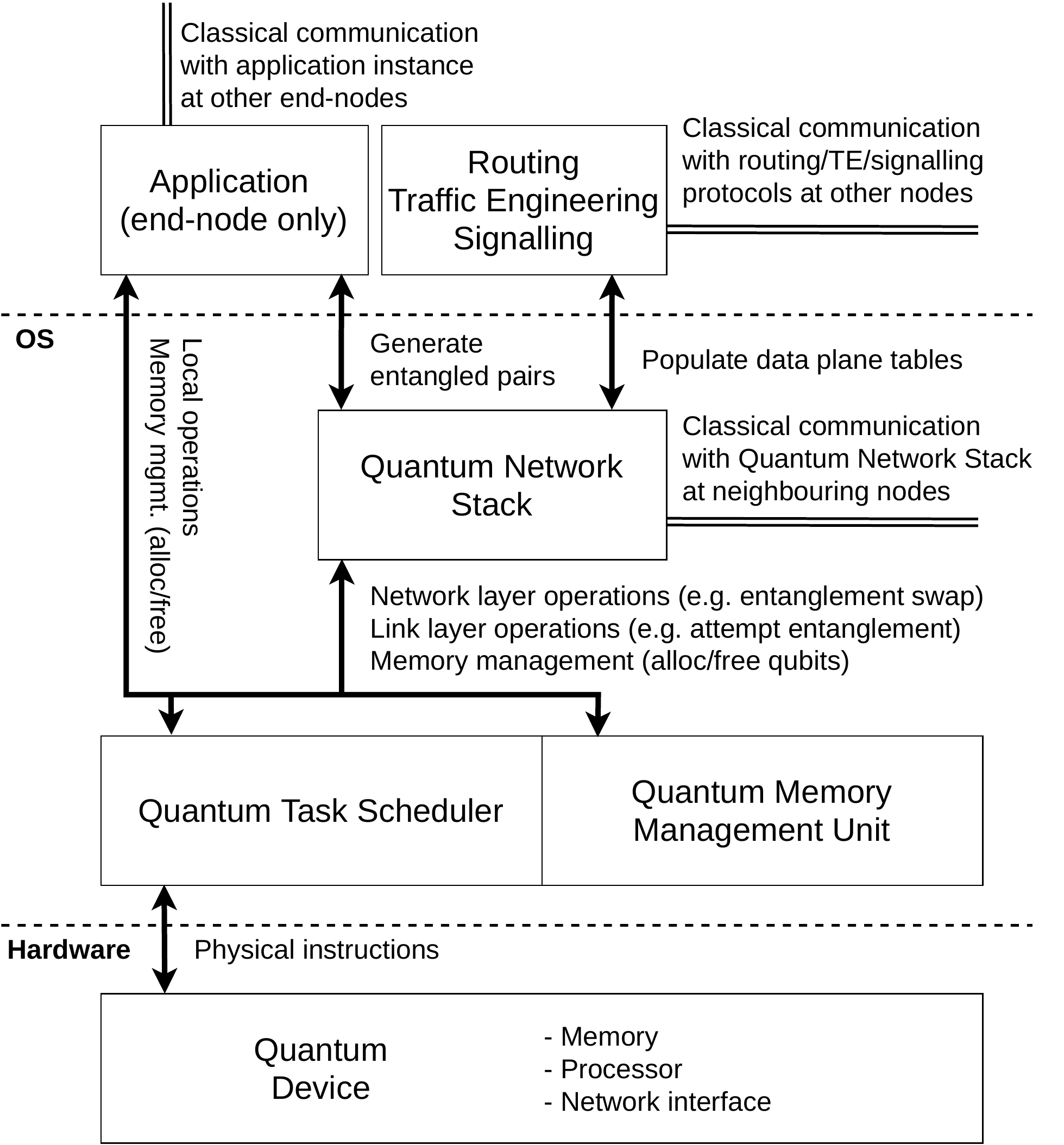}
  \caption{Local system components of a quantum node. The quantum memory,
    processor, and network interface are all one hardware component on current
    platforms. Local gate and network operations are performed on qubits in the
    main memory. Management and arbitration of local hardware resources belong
    to local operating system components such as a quantum task scheduler and a
    quantum memory management unit. }\label{fig:system}
\end{figure}

We first define the high-level architecture of a quantum node, shown in
\fref{fig:system}. The network stack is expected to be part of a local
operating system (OS). The stack is responsible for managing operations
relating to the generation of entangled pairs which it executes with the help
of local OS services.

Upon receiving an entanglement request (from an application or an upstream
node) the network stack will need to do two things: (i) coordinate with
neighbouring quantum nodes and (ii) issue local instructions to generate
entangled link-pairs and perform entanglement swaps. The network stack
coordinates with its neighbours by exchanging classical messages (all nodes are
connected classically, see \fref{fig:qnetwork}). Just like in classical
networks, certain tasks such as path computation happen outside of the network
stack itself. These tasks are delegated to other protocols which communicate
their decisions to the local network stack by means of populating relevant data
plane structures. Additionally, the network stack will have to issue
instructions to the local quantum device in order to generate link-pairs and
perform entanglement swaps. In currently available hardware, unlike in
classical devices, there is no distinction between the processor and the
network interface and they both operate directly on qubits in the main memory.
Though, in general, they are not able to operate on any arbitrary qubit on the
device. The precise nature of these limitations strongly depends on the
hardware implementation, but at a high level the qubits are split into
\emph{communication qubits}, those that can participate in networked
operations, and \emph{storage qubits}, those that can store quantum information
but cannot be used for entanglement generation~\cite{dahlberg2019link}. The
network stack relies on other OS components such as a quantum task scheduler
and a quantum memory manager for arbitrating access to hardware.

\section{The Quantum Network Layer}

\subsection{Use Cases}\label{sec:use}

Currently, no quantum networks exist so it is impossible to derive any use
cases based on real usage statistics. However, Ref.~\cite{dahlberg2019link}
identifies two categories of use cases that represent application demands of
quantum application protocols known to date: ``measure directly'' and ``create
and keep''.

\textbf{Measure directly} Applications in this category are characterised by
the fact that they consume the delivered pairs (by measuring them) as soon as
they are available and do not store them. Therefore, they can tolerate
fluctuations in the rate of delivery as the qubits never sit idly in memory
where they would decohere. This use case is relevant for applications that use
the entangled pairs to produce stronger than classical correlations such as
QKD~\cite{ekert1991quantum}, secure identification~\cite{damgaard2007secure},
other two-party cryptographic protocols~\cite{aharonov2000quantum,
  chailloux2011optimal, damgaard2008cryptography, ribeiro2015tight,
  wehner2008cryptography}, and other applications in the prepare-and-measure
stage of quantum networks~\cite{wehner2018quantum}.

\textbf{Create and keep} Applications in this category are characterised by
their need for storage, possibly of multiple entangled pairs simultaneously.
This use case is relevant for applications that may want to send qubits
deterministically (via teleportation), perform joint operations on multiple
qubits, or perform operations that depend on back and forth communication with
another node. Due to decoherence, these applications cannot tolerate large
delays between successive pairs. Examples of such applications include
sensing~\cite{gottesman2012longer}, metrology~\cite{komar2014quantum}, and
quantum distributed systems~\cite{ben2005fast, denchev2008distributed}.

\subsection{Service Delivered to Higher Layers}\label{sec:service}

Here, we explain the key aspects of the quantum network layer service delivered
to the higher layers.

\textbf{Entangled pair identifier} Logically, the network delivers an entangled
pair. Physically, the network delivers one entangled qubit to each of the two
end-nodes. This means that the network must track the entanglement swaps that
connect the individual link-pairs into a long-range pair such that at the end
it can identify which qubits at the end-nodes belong to the same pair. When
delivering the qubits, it provides this by means of an entangled pair
identifier.

\textbf{Entangled pair state} Entangled pairs come in four variants called the
Bell states. They are all equally usable, but the recipient must know which one
it has received. Due to the fundamental randomness of quantum mechanics, the
state of each pair produced by entanglement swaps is not known a priori, but is
revealed to the swapping node upon the swap's completion. The network must
collect these announcements, infer the state, and deliver this information to
the application.

\textbf{Class of service: fidelity} States do not have to be perfect to be
usable as long as they are above an application-specific threshold. Since more
time is needed to produce better states, applications can sacrifice fidelity in
exchange for higher rates (or vice-versa). Therefore, the user must specify a
minimum fidelity threshold, $F$, on each request. The network then attempts to
deliver these states. A strict guarantee is not required, because end-to-end
quantum security proofs do not rely on a trustworthy source of entanglement.

\textbf{Class of service: time}\label{sec:time} The application must be able to
quantify its desired fidelity-vs-rate trade-off, especially in light of the use
cases described in Sec.~\ref{sec:use}. For the ``measure directly'' use case,
the application can specify its requirement as either (i) $N$ pairs by deadline
$T$ or (ii) a rate of $R$ pairs per unit time. For the ``create and keep'' use
case the application specifies that it requires $N$ pairs by deadline $T$ such
that the last pair is delivered at most $\Delta t$ after the first. In both
cases $T$ may be set to zero to indicate no deadline.

\subsection{Network Layer Architecture}

Delivering the full network layer service cannot be accomplished with one
protocol alone. Instead, we envisage a situation similar to the one that exists
in classical networks where a variety of different services are built from
simpler building blocks such as the IP datagram or MPLS virtual circuits. In
this paper, we propose a \emph{quantum data plane} protocol that aims to
provide such a building block for quantum networks. However, our protocol
requires support from at least two external services: a signalling protocol and
a routing protocol. In this paper we only propose a quantum data plane
protocol, but we first outline the roles of the supporting protocols.

\textbf{Routing protocol} Before any end-to-end entangled pair can be generated
the optimal path must be determined. Just like in a classical network this is
expected to be done by a separate routing protocol. However, routing in quantum
networks is more complicated because it must compute the paths not only based
on path length, cost, and throughput, but it must also take into account the
desired end-to-end fidelity. Higher fidelity paths will require links that can
produce higher fidelity link-pairs and nodes with longer lasting memories.
Furthermore, higher fidelity link-pairs require more time to produce which must
be taken into account when determining available bandwidth. Routing algorithms
for quantum networks are an emerging field of
study~\cite{chakraborty2019distributed, chakraborty2020polynomial,
  caleffi2017optimal, gyongyosi2018decentralized, van2013path,
  schoute2016shortcuts, imre2012advanced, gyongyosi2017entanglement,
  li2020effective, shi2020concurrent}.

\textbf{Signalling Protocol} Our protocol is connection-oriented. It requires a
fixed path, called a virtual circuit, to be established between the end-nodes
prior to its operation. Installing virtual circuits will be the task of a
signalling protocol. This is similar to how RSVP-TE is used to install MPLS
virtual circuits in classical networks. However, allocating a path with
sufficient resources is not enough. In a quantum network the upstream and
downstream links at each node must generate their link-pairs sufficiently close
in time that they do not decohere before swapping. The routing component is
responsible for choosing a path based on available resources, but does not
decide how to use them. On the other hand, the quantum data plane protocol's
worldview will be limited to that of a single virtual circuit. We propose that
the signalling protocol is best suited to the task of schedule management. It
is an open question how best to perform scheduling at a quantum
node~\cite{vardoyan2019performance, vardoyan2019capacity}. In early-stage
network this synchronisation will have to be very precise and may need to
allocate dedicated time bins to each circuit.

These protocols can be implemented in a distributed or centralised fashion.
Researchers have considered both distributed~\cite{chakraborty2019distributed}
as well as centralised routing protocols~\cite{chakraborty2020polynomial,
  caleffi2017optimal, van2013path, li2020effective} in quantum networks. Our
design does not assume either architecture.

\subsection{Quantum Data Plane Protocol}\label{sec:tasks}

In analogy to classical networks, where the task of delivering connectivity
once all state has been installed is the responsibility of a data plane
protocol, in this paper we propose a \emph{quantum data plane} protocol. We
define the quantum data plane protocol as the component that is responsible for
coordinating the generation of link-level entanglement and the subsequent
entanglement swapping along a path between two distant nodes while minimising
the losses due to decoherence and compensating for the losses that do happen.
The focus of our work are mechanisms for the quantum data plane, that is, local
quantum operations and the classical messaging to coordinate these operations.
It is important to note that we do include classical message exchange that is
necessary to coordinate quantum operations in the definition of the quantum
data plane. However, it is not within the quantum data plane's domain to
perform any resource management, routing, or any other long-term state
management. Therefore, in this work we will assume the existence of suitable
routing and signalling protocols and focus on defining what information we
expect them to provide to the quantum data plane. This is in contrast to
existing works on quantum routing which focus on control plane aspects and
algorithms while working with an abstract model of the data plane.
Nevertheless, the quantum data plane protocol is expected to participate in
policing and shaping of the traffic to meet the use case requirements outlined
in Sec.~\ref{sec:use}. We expect the such a protocol to have three tasks:

\textbf{Link-pair generation management} To create a long-distance pair,
link-pairs must be first generated along the entire path. The network layer is
not expected to manage the physical process directly, but instead will rely on
a link layer protocol~\cite{dahlberg2019link} to deliver these link-pairs as
per the quantum network stack design shown in \fref{fig:stack}. However, it is
the network layer's responsibility to manage the link layer service at each
node along the path such that a sufficient amount of link-pairs of suitable
fidelity are produced.

\textbf{Entanglement swapping and tracking} Once the link-pairs are generated,
the repeaters must perform entanglement swaps to create long-distance entangled
pairs. In addition to performing the physical operation, the protocol must also
track the swaps that were involved in producing each end-to-end pair. This is
done for two reasons (outlined in Sec.~\ref{sec:service}): to correctly
identify which qubits that belong to the same end-to-end pair and which Bell
state they are in. Therefore, the network protocol needs a mechanism to collect
the entanglement swap outcomes and deliver them to the end-nodes so that the
final Bell state of the end-to-end pair can be inferred and delivered to the
recipient.

\textbf{Quality of service management} Whilst the quantum data plane protocol
cannot guarantee the quality of service on its own, it is expected to provide
basic mechanisms that will allow the supporting protocols to achieve this goal.
This includes at least (i) confidence that the fidelity is above the threshold,
(ii) policing requests by rejecting any that cannot be fulfilled, and (iii)
shaping traffic by delaying requests that can be fulfilled later.

\subsection{The Link Layer Service}\label{sec:ll}

\begin{figure}[t!]
  \centering
  \includegraphics[width=0.9\linewidth]{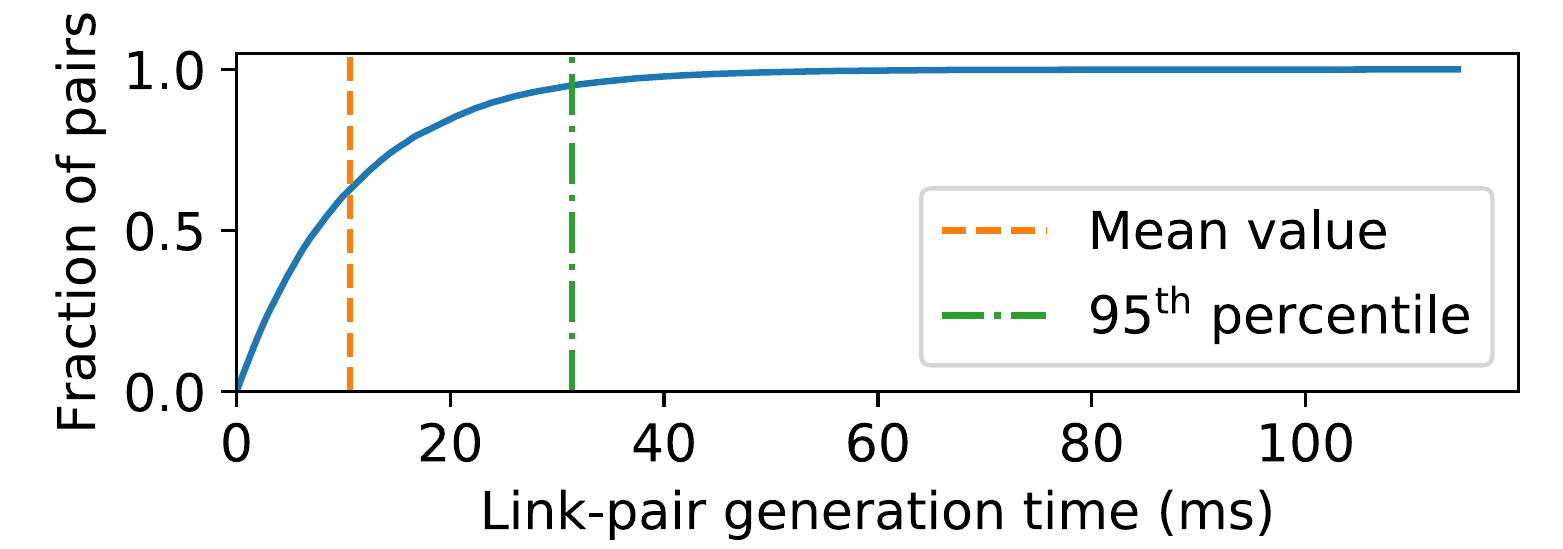}
  \caption{The cumulative distribution function for the time taken to generate
    a link-pair of fidelity 0.95 over a 2~m long fibre with the same hardware
    parameters as used in Sec.~\ref{sec:eval}. The y-axis denotes the fraction
    of pairs generated in less than the time indicated on the x-axis. We see
    that on average we have to wait 10~ms and that 95\% of link-pairs are
    generated within 30~ms.}\label{fig:delay}
\end{figure}

The link layer protocol interacts with the physical layer and exposes a
meaningful link entanglement generation service to the network layer. It is
meaningful in the sense that it is responsible for batching and multiplexing
entanglement attempts across a link in order to either deliver an entangled
pair to the higher layer with suitable identifiers or notify it of failure.
Since the probability of success on each entanglement generation attempt is
generally low, the link layer is expected to include a retry mechanism to
increase reliability. \fref{fig:delay} shows how long it takes to create a
link-pair.

A single link layer request is simply an asynchronous request made at one end
of the link which returns entangled qubits at both ends. Our network protocol
requires four properties from the link layer. (i) A link-unique request
identifier can be assigned to each link layer request. This identifier must
accompany all qubits delivered as part of this request at both ends --- this
allows the network protocol to coordinate its own actions across a link
(Purpose ID in Ref.~\cite{dahlberg2019link}). (ii) An identifier must be
assigned to each pair that uniquely identifies it within the particular link
layer request --- the network layer needs this for entanglement tracking
purposes (Entanglement ID in Ref.~\cite{dahlberg2019link}). (iii) The link
layer must inform the network layer which of the four Bell states the qubits
are delivered --- this information is needed for entanglement tracking in order
to infer the final state of the end-to-end pair. (iv) The caller must be able
to specify relevant quality of service parameters: minimum fidelity and time
restrictions --- this allows the network protocol to fine-tune its own quality
of service properties.

\section{Quantum Network Protocol}\label{sec:protocol}

\subsection{Protocol Design}\label{sec:design}

Here, we present the main design aspects of our quantum data plane network
protocol, the Quantum Network Protocol (QNP). A more detailed description is
available online~\cite{qnpspec}.

\textbf{Principle of operation} The QNP becomes operational once a virtual
circuit (VC) is installed into the network by the signalling protocol using the
path provided by the routing protocol. A VC is defined as a fixed path between
two end-nodes with the necessary data plane state installed into the local
network stack data structures. The circuit is directed with a head-end node at
the upstream end and a tail-end node at the downstream end. It is up to the
signalling protocol to determine which direction is upstream and which is
downstream. Whilst the entangled pairs are directionless this distinction is
used to give upstream nodes the right to initiate pair-wise activities, such as
link-pair generation.

The QNP starts when a request is received at the head-end node (for simplicity
we currently require the tail-end node to forward user requests to the head-end
node). This triggers a FORWARD message sent downstream towards the tail-end
node initiating link-pair generation for this particular VC on each link along
the path. Once two link-pairs are generated at the same intermediate node, one
on the upstream and one on the downstream link, an entanglement swap is
immediately performed (without any further classical communication). The swap
outcomes are collected by two TRACK messages, one going downstream and one
upstream. Once the TRACK messages reach the end-nodes the pair is delivered to
the application. Some applications can consume the qubits before the TRACK
messages arrive which we discuss later.

\textbf{Virtual circuits} The central property of our protocol is that it is
connection-oriented. That is, a connection, in the form of a VC installed by
the signalling protocol, must be established prior to the QNP's operation. This
decision is driven by the fact that link-pair generation and entanglement
swapping are parallelisable. Link-pairs themselves are completely independent
of each other until they are connected via an entanglement swap so they can all
be generated at the same time. Furthermore, the order in which the entanglement
swaps are executed also does not matter. VCs enable parallelisation as they
allow us to dedicate resources on each link along the path for a particular
end-to-end connection. Since link-pair generation is not necessarily a fast
process (rates in laboratory setups are of the order of few tens of
Hz~\cite{humphreys2018deterministic}) this is a significant performance
optimisation. Short memory lifetimes further compound the benefits of
parallelisation as it allows the protocol to minimise the decoherence
experienced by the qubits --- they will not have to wait as long for a matching
qubit to become available for swapping.

VCs are installed by the signalling protocol. It achieves this in a similar
manner to MPLS\@: by assigning a link-unique label, called the link-label, to
each link on the path of the circuit. The network protocol then uses this label
as its request identifier when issuing requests to the link layer service.
Entanglement swaps are performed as soon as pairs with labels for the same VC
are available on the upstream and downstream links.

It is worth noting some works~\cite{pant2019routing, shi2020concurrent} on
routing entanglement in quantum networks assume a different model for the
quantum data plane. Instead, they build upon an abstract model of the network
whereby some subset of (or all) links in the network attempt to generate
entanglement in pre-defined time slots. Swapping decisions are then made by
each node based on their knowledge of global topology combined with information
about which of the nearby links succeeded in generating entanglement in that
time slot. These quantum data plane models show good performance when used in
conjunction with the aforementioned routing protocols. However, they rely on
networks that are sufficiently big that they are able to support multiple paths
between the relevant source and destination pairs which will not be the case
for near-term deployments. Our quantum data plane protocol does not have this
problem as it is designed to be operational on single paths. However, as our
protocol is inspired by MPLS VCs, generalising it to multipath scenarios (for
redundancy or bandwidth purposes) will be straightforward at which point it may
also be used with multipath entanglement routing protocols.

\textbf{Swap records} As explained in Sec.~\ref{sec:tasks} the protocol must
track the entanglement in addition to performing entanglement swaps. That is,
it must (i) correctly identify which qubits at the end-nodes are part of the
same entangled pair and (ii) collect all the entanglement swap results to infer
the final Bell state of the end-to-end pair. For this reason, as soon as an
entanglement swap completes, a temporary swap record is logged at the node.
This record must contain the following information: (i) the link-unique
identifiers (Entanglement ID) for the two pairs involved in the swap and (ii)
the two-bit output of the entanglement swap. Provided the Bell states of the
input pairs are known, the two-bit output uniquely identifies the Bell state of
the output pair which now spans between the two remote qubits of the two input
pairs (see \fref{fig:swap}).

\textbf{Lazy entanglement tracking} The swap records generated after every
entanglement swap must be collected and sent to the end-nodes so that they can
deliver the end-to-end pair with the correct identifier and Bell state
information. The QNP achieves this by sending an entanglement tracking message
(TRACK) from the head-end to the tail-end along the VC which collects the
records at each node it visits, waiting if a swap has not completed yet. A
similar message is sent in the reverse direction so that the head-end can also
receive this information.

An individual swap record is sufficient to identify the Bell state of the
output entangled pair provided the input Bell states are known. The problem is
that in a VC with multiple intermediate nodes where the ordering of the
entanglement swaps is not defined, the swapping nodes do not actually know what
the input states are (as other swaps along the VC may have already happened
changing the state of the input pairs) so they are unable to infer the output
state from their swap record on their own. However, we do not need the swapping
nodes to know this information --- it is only needed at the end-nodes. The
TRACK messages collect the swap records one by one from one end-node all the
way to the other end-node. As the ordering of the entanglement swaps does not
matter, logically, a TRACK message can be thought of as reconstructing the
final entangled pair state as if the swaps happened in the order it collects
these swap records. In this context, the TRACK message effectively carries
information about the input state for the next swap. When it collects a new
swap record, it uses the two-bit swap outcome contained within to infer the new
input state for the swap at the next node. When it reaches the final end-node
this ``next input state'' is actually the final entangled pair state. This
logical picture works for TRACK messages in both directions (upstream and
downstream) as the ordering of swaps does not matter.

We call this lazy entanglement tracking, because the protocol does not keep
track of any of the intermediate pairs created throughout the process. The
swaps do not necessarily happen in the order the TRACK messages collect the
records so they do not represent the intermediate states as they really
happened. The only pair the TRACK message is guaranteed to know the state of is
the final pair. This allows: (i) quantum operations to proceed regardless of
classical control messages being communicated and (ii) individual nodes to
discard decohered qubits (discussed later) without having to separately
communicate this with the rest of the VC\@.

The ability to do lazy entanglement tracking is an advantage of the
connection-oriented approach as opposed to a hop-by-hop model where each node
makes a swapping decision without any prior agreement. In that case it would be
necessary to keep track of all intermediate pairs in order to know what pair
will result from the next swap. This would introduce additional latency and
synchronisation issues as the protocol would need to constantly update its
entanglement information database. In the worst case this will block
entanglement swaps until the protocol completes synchronising this information
which is highly undesirable, especially in the presence of decoherence.

\textbf{Cutoff time} When memory lifetimes are short (as will be the case for
near-future hardware), it often happens that a qubit may decohere too much by
the time a suitable pair on another link is available. To counteract this, we
adopt the cutoff mechanism from repeater chain
protocols~\cite{khatri2019practical, rozpedek2018parameter, rozpedek2018near,
  li2020efficient}. The protocol discards qubits that have not been swapped,
but have reached a cutoff deadline. The tighter the deadline the less likely it
is that two links will be able to generate link-pairs at the same time, but
when they do manage to be generated within the cutoff window the qubits will
have suffered from less decoherence leading to a higher end-to-end fidelity.
Therefore, we allow the external routing protocols to specify the cutoff value
as well. These timeouts can then be distributed by the signalling protocol when
setting up the circuit.

When a qubit is discarded, the node must log a temporary discard record. When
an entanglement tracking message arrives at the node, it will check for the
discard record if it cannot find a swap record. If the discard record is
present, the tracking message will be sent back to its origin to notify that
end-node of the broken chain. The cutoff timer is not used at the end-nodes as
we found this to result in a window condition where one end-node delivers its
end of the pair to the application whilst the other end-node discards the other
qubit. Therefore, the end-nodes instead discard their qubits upon receipt of
this expiry notification.

\textbf{Policing and shaping} If circuits are used with a resource reservation
mechanism they will also be allocated a maximum end-to-end rate (EER),
i.e.~bandwidth. The routing protocol computes a path that can support a given
EER and the signalling protocol provides the head-end node with this EER value
so that the QNP can police (reject) and shape (delay) incoming requests. The
head-end node calculates a request's minimum EER, compares it to its available
bandwidth and decides if the request can be satisfied by the specified deadline
$T$. Our service definition from Sec.~\ref{sec:time} requires applications to
always specify their minimum rate in its request which we use as its minimum
EER (measure directly: $N/T$, $R$, or 0 if $T$ not set; create and keep:
$N/\Delta t$). Note that these checks are only made at the end-nodes and we do
not implement any further in-network mechanisms. It is the role of the resource
reservation protocols to ensure that network resources are not over-subscribed
as long as the end-nodes fulfil their part of the contract by policing and
shaping the incoming requests.

\textbf{Continuous link generation} Discarding qubits due to decoherence will
be the norm rather than the exception in early-stage networks. Therefore, an
efficient retry mechanism is necessary. For this reason, the quantum network
protocol simply requests the link layer service to produce a continuous stream
of pairs until the end-nodes signal the completion of the request. To allow the
link layer to multiplex requests from different circuits, the network layer
must provide some information about the desired rate. The link-pair rate (LPR)
must necessarily be higher than the EER as some link-pairs will be discarded
due to decoherence. The routing component will have calculated the necessary
LPRs for each link when determining which path can support a given
EER~\cite{chakraborty2020polynomial}. The QNP will request the maximum LPR on
each link unless the only active requests are rate-based (``measure directly''
requests that specify $R$) in which case it requests a suitable fraction of the
circuit's LPR based on the fraction of its EER that these requests need.

\textbf{Early delivery} For the ``measure directly'' use case the application
may benefit from acting on its entangled pair as soon as possible to minimise
decoherence. Some applications can start operating on the qubit at their
end-node before all entanglement swaps complete --- the effect will be
propagated with the swaps to the other end. Thus, they may choose to have the
QNP perform a measurement as soon as its end of the pair is available or have
it delivered before the protocol can confirm the pair's creation. In the case
of a measurement, the protocol simply withholds the result until the tracking
messages arrive so that only results from successful pairs are delivered. If
the pair was delivered early, the application must take over the responsibility
of handling any error messages such as notifications about discarded pairs. It
will also have to wait for the final entanglement tracking information of the
entangled pair to correctly post-process its results.

\textbf{Aggregation} Entangled pairs generated between the same two end-nodes
for the same fidelity threshold are, for application purposes,
indistinguishable. Therefore, the QNP may aggregate such requests onto the same
VC\@. Aggregation is an important feature of the protocol that enables
scalability, because (i) it reduces the amount of state the network needs to
manage by reducing the total number of circuits, and (ii) it improves resource
sharing at entanglement swapping nodes. To explain the second point, we note
that a repeater node may only swap two entangled pairs if they belong to the
same circuit. Without aggregation, a node would have to wait for two pairs
allocated to the same request before swapping. With aggregation the nodes do
not have to distinguish between individual requests if they share the same
VC\@.

Aggregation means that the VC does not keep track of any request information.
Therefore, demultiplexing, i.e.~assigning a VC's pairs to requests, must be
done by the end-nodes. There are many ways to do this. The QNP only requires
that the end-nodes agree on a method which can be negotiated when the VC is set
up. The end-nodes may use a distributed queue, have the head-end node make all
the decisions and communicate them on the TRACK messages, or use some other
algorithm. We do not specify the strategy as part of the protocol. However, we
do provide two mechanisms to aid in this task. (i) Epochs: an epoch is the set
of currently active requests. A new epoch is created (but does not activate)
whenever a request is received or completed. The head-end advances the active
epoch by setting the value of the next one on each TRACK message. Once the
entangled pair corresponding to that TRACK message is delivered the epoch
indicated by that message becomes active. (ii) TRACK messages carry information
about which request they were assigned to by the end-node that originated the
message. Due to the cutoff strategy, qubits along the VC may be suddenly
discarded which leads to window conditions where the end-nodes may not agree on
which request the pair was assigned to. This information allows the end-nodes
to perform a cross-check and discard such qubits if necessary (if a qubit was
not delivered early it may even be possible to reassign it).

\textbf{Routing table} To communicate all the routing decisions to the quantum
data plane protocol, we require a routing table entry at each node for each
VC\@. This entry must contain: (i) the next downstream node, (ii) the next
upstream node (TRACKs are also sent upstream), (iii) the downstream link-label,
(iv) the upstream link-label, (v) the downstream link minimum fidelity, (vi)
the downstream maximum LPR, and (vii) the circuit maximum EER\@. The fidelity
threshold for a link will be higher than the end-to-end fidelity to account for
losses due to entanglement swapping and decoherence. The nodes are also
provided with the circuit maximum EER so that the QNP can scale its LPR if the
VC's maximum EER is not required. We delegate the responsibility for choosing
the fidelity and LPR values to a routing protocol for two reasons: (i) choosing
them requires knowledge of the entire path --- the longer the path, the higher
must they be on each link to compensate for various losses --- and (ii) their
exact values depend on the hardware parameters of all the nodes and links on
the path of the VC\@.

It is worth noting that the LPR and link fidelity values do not have to be
identical for every link along the path of a particular VC\@. In fact, this is
likely to be the case in heterogeneous networks where the different links have
different rate-fidelity trade-offs and the nodes have different decoherence
rates. Assuming isotropic noise (i.e.~the worst case scenario) so that the
entangled pairs can be expressed as Werner states~\cite{werner1989quantum} it
can be shown that the fidelity, $F^\prime$, of an entangled pair produced by
combining two pairs with fidelities $F_1$ and $F_2$ is given by
\begin{equation*}
F^\prime = F_1 F_2 + \frac {\left( 1 - F_1 \right) \left( 1 - F_2 \right)} {3}.
\end{equation*}
This expression is associative and thus variations in link fidelity do not
affect our key assumption that entanglement swaps can be performed in any
order. Therefore, in heterogeneous networks it is conceivable that the fidelity
is ``budgeted'' differently across the different links as necessary to improve
the end-to-end rates.

\textbf{Fidelity test rounds} It is physically impossible for the protocol to
peek or measure the delivered pairs to evaluate their fidelity. However, we
need a mechanism to provide some confidence that the states delivered to the
application are above the fidelity threshold. We apply the same method as
proposed in Ref.~\cite{dahlberg2019link} for individual links, but instead test
end-to-end pairs. In summary, the method relies on creating a number of pairs
as test rounds which are then measured (and thus consumed). The statistics of
the measurement outcomes can be used to estimate the fidelity of the non-test
pairs.

\textbf{Classical communication and link reliability} The protocol requires
that all its control messages are transmitted reliably and in order. It is
designed to not depend strongly on the classical messaging latency so that we
may simply rely on a transport protocol to provide these guarantees (e.g.~TCP
or QUIC). Every VC establishes its own transport connection between every pair
of nodes along its path for this purpose. The transport's liveness mechanism
can then be used to monitor the classical channel liveness and tear down the VC
if the connection goes down. The quantum link layer is also expected to support
a liveness check mechanism (Ref.~\cite{dahlberg2019link} does in the form of
fidelity testing rounds). If a circuit goes down due to loss of connectivity,
the protocol aborts all requests and notifies applications of the failure.

\subsection{Example Sequence}

\begin{figure}[t!]
  \centering
  \includegraphics[width=\linewidth]{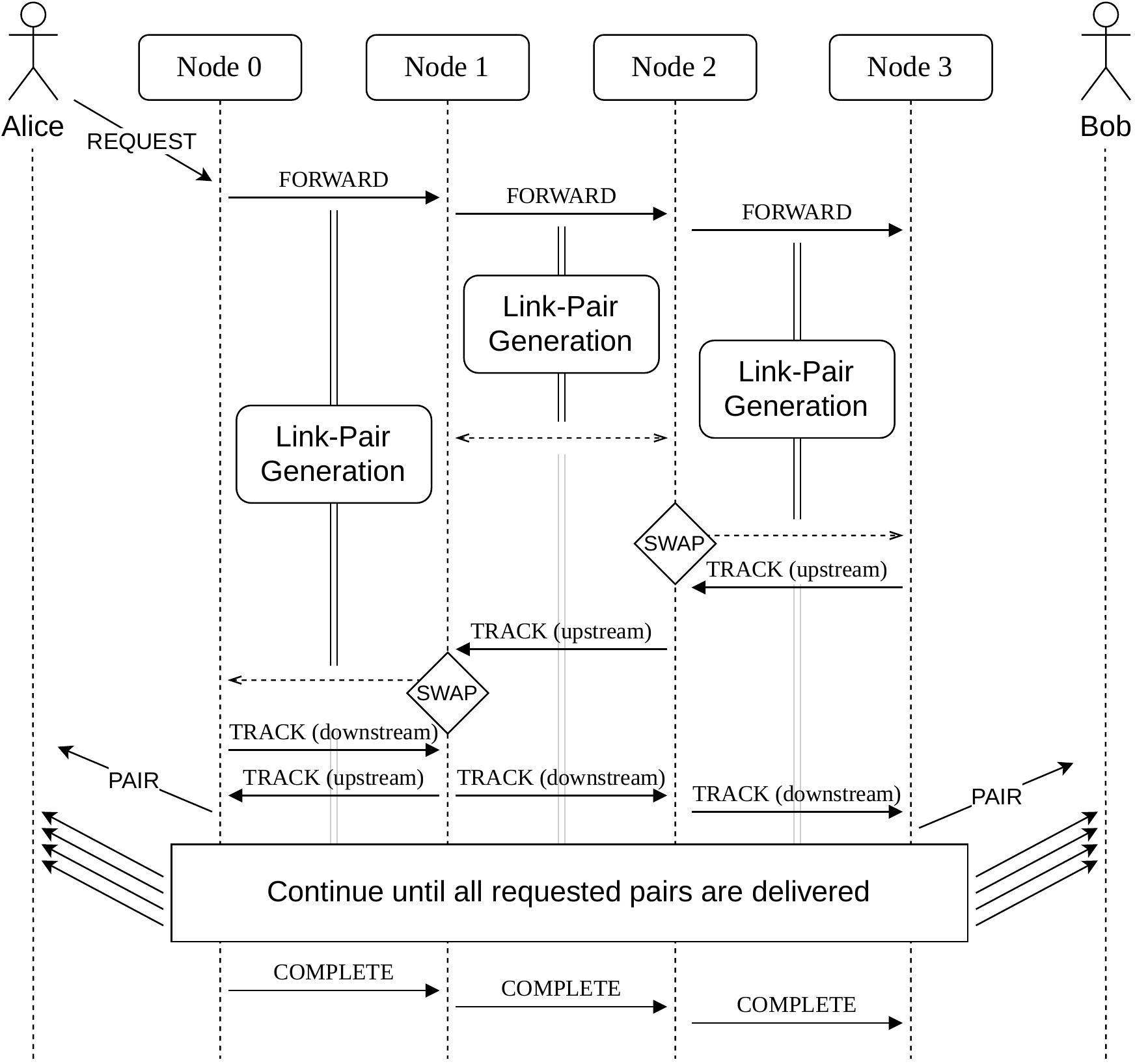}
  \caption{Example sequence of the QNP.}\label{fig:qnp}
\end{figure}

\fref{fig:qnp} illustrates a sequence diagram of a sample flow. Upon receiving
a request, a FORWARD message is sent along the VC to initiate link-pair
generation. Entanglement swaps execute as soon as an upstream and downstream
pair are available for the same circuit and a swap record is generated upon its
completion. Each end-node initiates a TRACK message as soon as their link-pairs
are available. The TRACK messages proceed along the circuit collecting swap
records, waiting for the corresponding pair's swap to complete if necessary.
Once the TRACK messages arrive at the destination end-nodes, the final
identifier and Bell state information are delivered together with the qubit
itself, if not delivered early. Once all pairs are delivered, a COMPLETE
message is sent along the circuit to terminate/update the link layer requests.

\subsection{Entanglement Distillation}\label{sec:distil}

Entanglement distillation is a process through which two or more imperfect
pairs are consumed to produce a higher fidelity pair with some finite
probability~\cite{dur2007entanglement, kalb2017entanglement}. However, because
entanglement distillation has higher hardware requirements than entanglement
swapping, it is not the solution to extremely fast decoherence. Nevertheless,
it will be necessary to overcome poor quality links and the fundamental loss of
fidelity due to entanglement swapping which ultimately limits the achievable
path length.

We decided not to incorporate distillation into the protocol at this stage of
development of quantum networks, because it is still an open research question
as to what the right distillation strategy is: should distillation happen as
soon as link-pairs are generated, after every swap, after $N$ swaps, at the
ends only, etc. Furthermore, there are many different methods available for
performing distillation, each with its own
trade-offs~\cite{rozpedek2018optimizing}. However, the QNP was designed to be
used as a building block for more complex quantum network services and
entanglement distillation offers a particularly interesting example of such a
service. Therefore, we instead illustrate how distillation could be implemented
on top of our protocol.

To implement distillation using the QNP we rely on the observation that this
process consumes two or more entangled pairs between the same pair of nodes.
Therefore, one can implement distillation in a layered fashion. We run the
network protocol between a pair of intermediate nodes which deliver entangled
pairs to a distillation module. Once distilled, the module passes the higher
fidelity pair to another circuit that only runs between the distillation
end-points and that sees all the nodes in between as one virtual link. This
proposal is similar to some of the early quantum network architecture
proposals~\cite{van2013designing}.

\section{Evaluation}\label{sec:eval}

To evaluate the performance of the QNP we have implemented it on top of a
purpose-built discrete event simulator for quantum networks called NetSquid
(Python/C++)~\cite{netsquid}. The simulator is responsible for the accurate
representation of the physical hardware including decoherence, propagation
delay, fibre losses, quantum gate operations and their time dependence. The
protocol itself is implemented in Python and runs on top of the link layer
implementation from Ref.~\cite{dahlberg2019link}.

As our work is focused on quantum data plane processes we keep the control
plane as simple as possible. For routing purposes we implement a rudimentary
algorithm that runs in a central controller and assumes all links and nodes are
identical. It calculates a network path together with link fidelities as a
function of end-to-end requirements by simulating the worst case scenario where
every link-pair is swapped just before its cutoff timer pops. The routing
information is installed by a source-routed signalling protocol. We also
implement a simple swapping and link scheduling algorithm. Links function
independently of each other and schedule requests using a weighted round-robin
scheme where the number of pairs generated for a particular VC is proportional
to its LPR and inversely proportional to the average time per pair. This
mechanism ensures that: (i) circuits get an equal share of the link's time
regardless of fidelity (higher fidelity VCs need more time to achieve the same
rate), (ii) when under-subscribed the excess capacity is distributed
proportionally to demand, (iii) when over-subscribed the available capacity is
distributed proportionally to demand. At each node, each VC will maintain two
logical queues of link-pairs (upstream and downstream) ready for swapping. Note
that these queues are only logical and they must all share a limited number of
physical qubits for storage purposes --- we do not pre-allocate qubits to
particular VCs. For queuing entanglement swaps we employ the first in, first
out strategy with the caveat that qubits may expire due to the cutoff timer.
That is, entanglement swaps always prefer the oldest unexpired upstream and
downstream pairs that correspond to the same VC\@. We do not perform any
resource management (all VCs are admitted regardless of available bandwidth) as
it is an open research question beyond the scope of this paper. Instead, we
examine the protocol's performance under different loads and draw conclusions
as to how resources could be managed.

\begin{figure}[t!]
  \centering
  \includegraphics[width=0.9\linewidth]{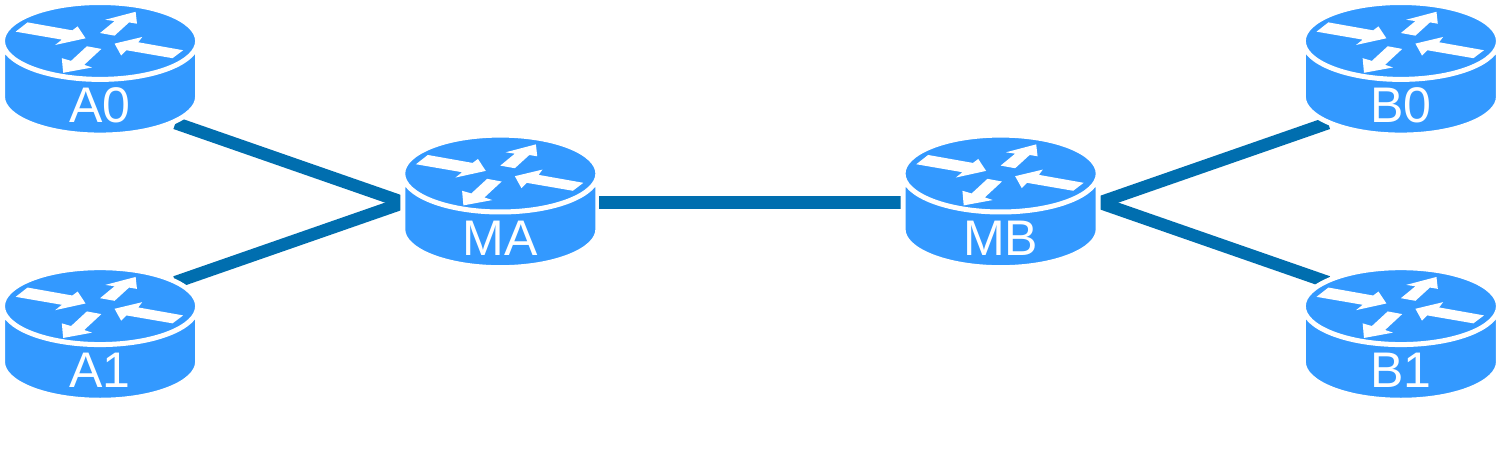}
  \caption{The evaluation topology. MA-MB is a bottleneck link between the A
    nodes and the B nodes. All links consist of a quantum and a classical
    channel.}\label{fig:topo}
\end{figure}

For the evaluation we consider the topology shown in \fref{fig:topo} which has
six nodes in total, four of which we use as end-nodes (A0, A1, B0, B1), and
with one bottleneck link (MA-MB). The dumbbell topology is complex enough that
it is functionally beyond the capabilities of repeater chain protocols and
requires the ability to merge and split entangled pair flows. At the same time
it is simple enough that the control plane does not have to make any difficult
routing decisions allowing us to focus our evaluation on the quantum data plane
processes. Our simulation is based on a simplified model of nitrogen vacancy
centre repeater platform~\cite{abobeih2018one, bradley201910,
  kalb2017entanglement, taminiau2014universal, reiserer2016robust,
  cramer2016repeated, humphreys2018deterministic, riedel2017deterministic,
  zaske2012visible}. We simplify the model by allowing for arbitrary quantum
gates and increasing the number of communication qubits from one per node to
two per link (not shared between links). The exact hardware parameters used are
listed in Appendix~\ref{app:hw}. For the entire evaluation except for
Sec.~\ref{sec:near-future} we consider parameters that are slightly better than
currently achievable. The parameters were chosen to produce higher fidelities,
but retain rates comparable to current hardware. The links between the nodes
are 2~m in length and we do not convert the photons to telecom wavelength. We
set the cutoff timeout to the time it takes a link-pair to lose approximately
1.5\% of its initial fidelity. We run each simulation 100 times and calculate
the average values of the quantities. Error bars are not shown as they are
comparable to, or smaller than, the plot markers, unless stated otherwise.

\subsection{Throughput and Latency}

\begin{figure*}[t!]
  \centering
  \includegraphics[width=0.9\textwidth]{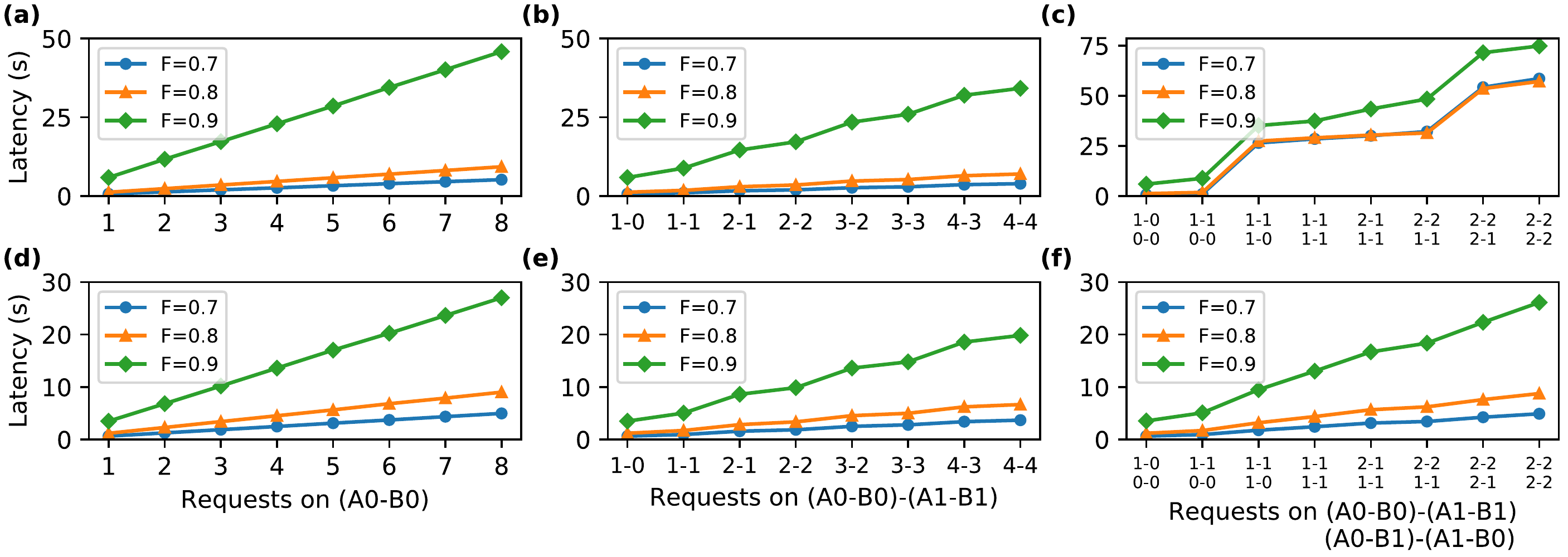}
  \caption{ Average latency of requests on the A0-B0 circuit when 1--8
    simultaneous requests, each for 100 pairs, are issued across (a,d) 1
    circuit (A0-B0), (b,e) 2 circuits (A0-B0, A1-B1), and (c,f) 4 circuits
    (A0-B0, A1-B1, A0-B1, A1-B0). We consider a long (a-c) and short (d-f)
    cutoff time (see main text). Linear growth in (a,b,d,e) shows that circuits are
    efficiently shared across multiple requests. A shorter cutoff allows the
    routing algorithm to use a tighter bound on the decoherence and thus to
    relax the fidelity requirements on each link improving their rates. In (c)
    the 4 circuits struggle to share the bottleneck link when the cutoff time
    is long. Our scheduling algorithm is too simple and often generates pairs
    which do not have a matching pair on the same circuit on another link.
    Reducing the cutoff time (f) alleviates this problem as pairs that cannot
    be swapped are discarded faster. }\label{fig:lat}
\end{figure*}

To gain some intuition about the protocol, before we study the effect of major
decoherence, we evaluate it on devices with long memory lifetimes of one minute
(current record on nitrogen vacancy platform not connected to a
network~\cite{bradley201910}). We first investigate how the protocol shares
resources in the network when multiple VCs have to compete for resources. We
investigate the end-to-end latency of multiple requests issued across multiple
circuits that all share the MA-MB bottleneck link. We simultaneously issue
between 1--8 requests for 100 pairs each split across up to four circuits. We
consider three scenarios: one circuit only (A0-B0), two circuits (A0-B0,
A1-B1), and four circuits (A0-B0, A1-B1, A0-B1, A1-B0). We vary two parameters:
the end-to-end fidelity and the cutoff time. Normally we set the cutoff time to
a value determined by the memory lifetime, but here we are using a relatively
long-lived memory so we will also consider a ``shorter cutoff'' set to the time
it takes for a link to have a 0.85 probability of generating a link-pair (see
\fref{fig:delay}). The requests are equally distributed across the circuits in
a round-robin manner. For example, in the four circuit scenario with six
requests, the circuit A0-B0 handles the 1$^\text{st}$ and 5$^\text{th}$
requests, circuit A1-B1: the 2$^\text{nd}$ and 6$^\text{th}$, A0-B1: the
3$^\text{rd}$, and A1-B0: the 4$^\text{th}$. All VCs are set up with the same
max-LPR on the bottleneck link so they all get the same share of that link's
time regardless of how many requests they carry. The average end-to-end request
latency of requests issued on the A0-B0 circuit are shown in \fref{fig:lat}. It
is immediately obvious that higher end-to-end fidelity pairs take longer to
generate.

In \fref{fig:lat}~(a,b,d,e), we also see that when requests are split across up
to two circuits, the latency scales linearly with the number of requests across
the bottleneck link. However, \fref{fig:lat}c shows that the network struggles
to multiplex four circuits (a ``quantum congestion collapse''). Our scheduling
algorithm is too simple: it assumes the links are independent, but they are
not. A pair on an upstream link must wait for a pair on the downstream link to
be generated for the same VC\@. Therefore, with four circuits and only two
qubits per link, it can happen that no VC has matching pairs in their upstream
and downstream queues and with no free qubits in the quantum memory the links
cannot generate more pairs. The requests complete, because eventually the pairs
decohere and are discarded. This problem can be solved by either not admitting
this many circuits or by improving the scheduling and/or queuing at the nodes.
\fref{fig:lat}f shows that reducing the cutoff value (effectively modifying the
local scheduling strategy) alleviates the problem. A shorter cutoff improves
throughput as any pairs that are using up memory slots without swapping are
discarded sooner. Nevertheless, more research is required as to what the best
scheduling strategy might be. We also note that the 1- and 2-circuit cases
benefit from the shorter cutoff time. This is because a shorter cutoff allows
the routing algorithm to use a tighter bound on the time qubits spend idling
and as a result it can relax the fidelity requirements on each link leading to
improved rates.

\begin{figure}[t!]
  \centering
  \includegraphics[width=0.9\linewidth]{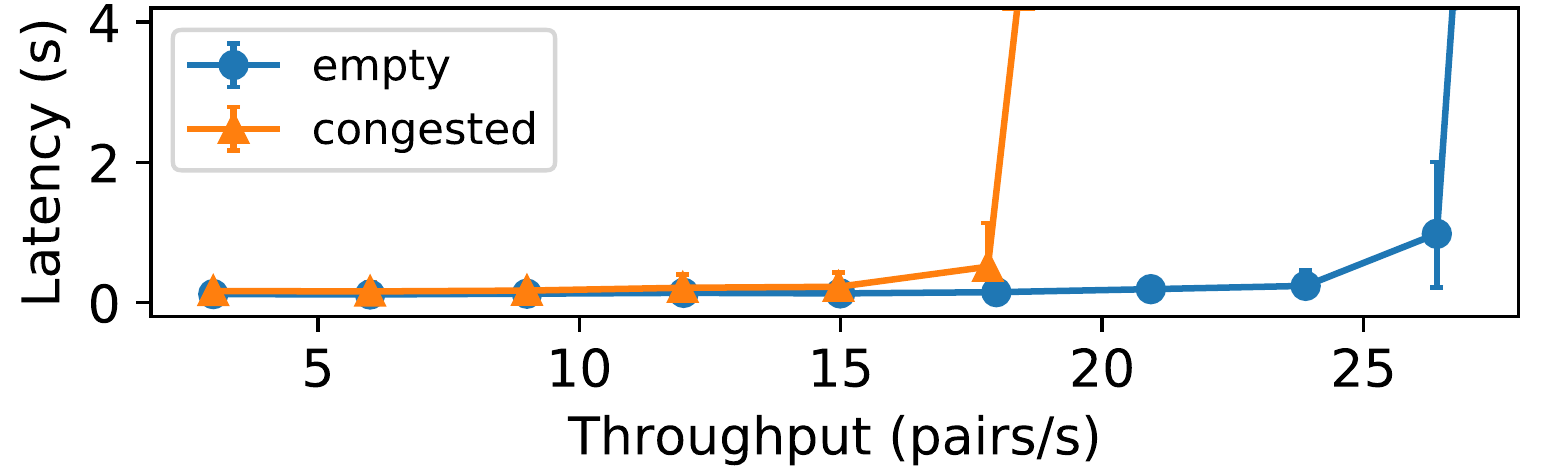}
  \caption{ Average latency vs.~throughput of A0-B0 circuit as we increase the
    rate of 3-pair requests over A0-B0. In the ``empty'' case, there is no
    other traffic in the network. In the ``congested'' case, there is a long
    running flow on A1-B1 at the same time competing for the bottleneck link.
    Error bars denote 5th and 95th percentile of the measured latency.
  }\label{fig:latthru}
\end{figure}

In the previous example, all requests were using their share of the bottleneck
to the fullest. To evaluate how request latency scales with throughput we issue
a series of smaller requests, each for three pairs, at an increasing frequency
at regular intervals. This time, we only consider two circuits: A0-B0 and A1-B1
and we use the shorter cutoff. We send the small requests over the A0-B0
circuit and measure their latency and the VC's throughput. We run this scenario
in an empty network (A1-B1 is idle) and in a congested network (A1-B1 is
constantly busy with a long running request). We run the simulations for 50
simulated seconds and measure the latency of requests issued after the 40~s
mark (a saturated equilibrium). \fref{fig:latthru} shows the average request
latency vs the VC throughput. The latency is constant until the link saturates.
The A0-B0 VC in the congested case saturates at more than half the value of the
empty case. Whilst counter-intuitive, this has a simple explanation: the MA-MB
link is shared by two circuits and thus generates pairs for each circuit slower
than the non-congested links. Therefore, the other links will have a higher
probability of having a pair ready for a swap by the time the MA-MB pair is
ready.

\subsection{Decoherence}

\begin{figure}[t!]
  \centering
  \includegraphics[width=0.9\linewidth]{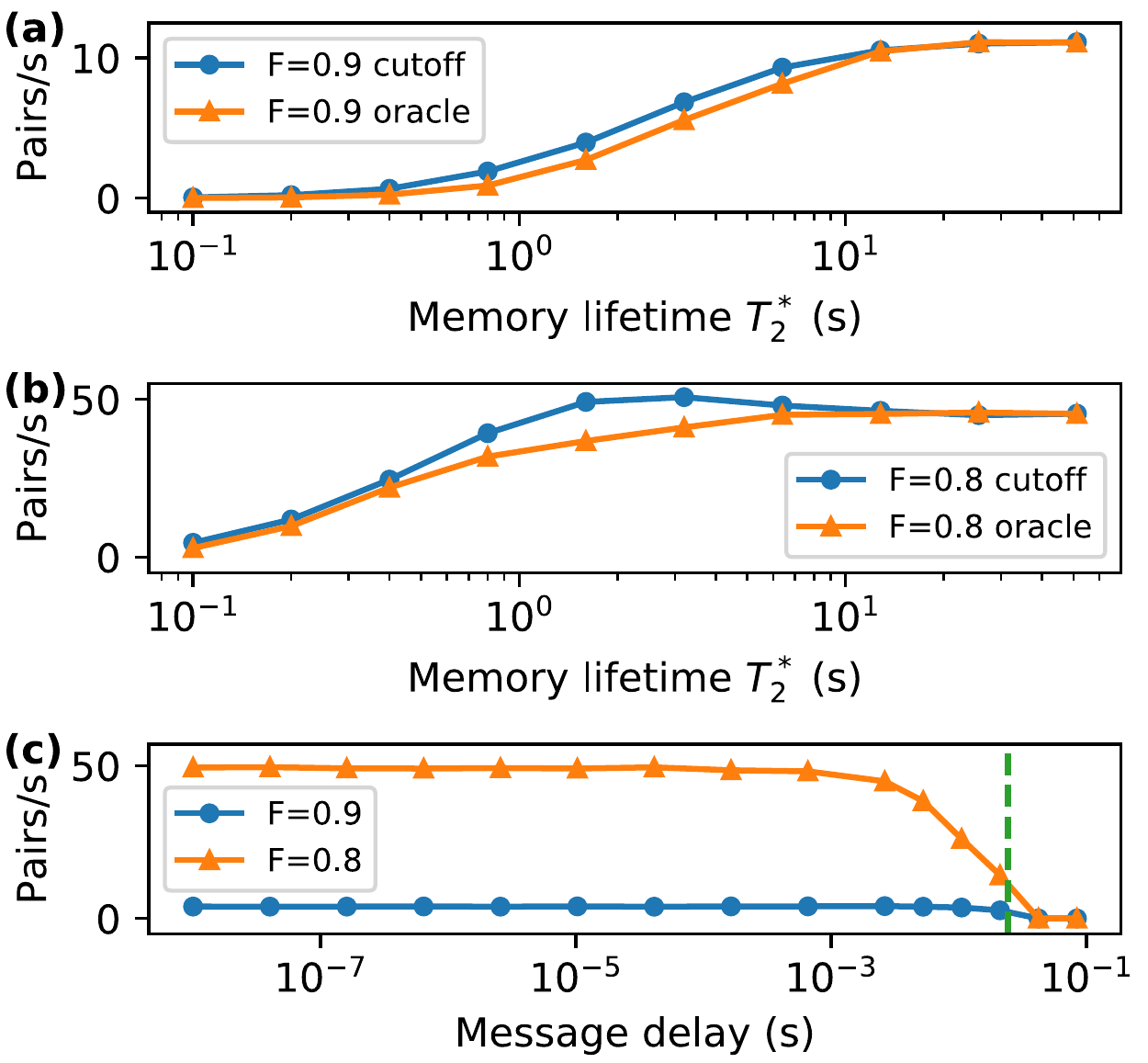}
  \caption{Robustness against decoherence. (a,b) show the effects of short
    memory lifetime on the throughput of the two competing circuits. Note that
    the F=0.9 with cutoff throughput becomes low, but not zero. (c) shows the
    effects of classical message delay (time from sending from one node to
    processing at next node). The dashed vertical line is the qubit cutoff
    value.}\label{fig:decoherence}
\end{figure}

We evaluate the two mechanisms for handling decoherence: the cutoff timer and
not forcing quantum operations to wait for control messages. Here, we evaluate
the protocol by running two circuits: A0-B0 for a fidelity of 0.9 and A1-B1 for
a fidelity of 0.8. We use different fidelity values for the two VCs as lower
fidelity requests suffer less from decoherence as the link-pairs are generated
faster and can tolerate longer idle times. We issue one long running request
for each circuit. The bottleneck link will round-robin between the two circuits
allocating the same amount of time to each. Since the 0.8 fidelity circuit
requires less time to generate each link-pair it will operate at a faster rate.
We stop the simulation after 20~s of simulated time and calculate the
throughput.

\textbf{Cutoff timer} \fref{fig:decoherence}~(a,b) shows the throughput of both
VCs against the memory lifetime parameterised by $T_2^*$, the dephasing time of
a qubit~\cite{nielsen2000quantum}. We see that as the memory lifetime decreases
so does the throughput due to an increased rate of qubits being discarded.
Higher fidelity VCs are affected more significantly as it takes longer to
generate the link-pairs and thus they have a smaller window for swapping. In
both cases we compare the performance of the protocol to a simpler protocol
which instead of using a cutoff in the network discards end-to-end pairs that
are below our required fidelity threshold. However, knowing which pairs are
below the fidelity threshold is highly non-trivial as it is not possible to
simply read it out from a pair. It would require a fidelity tracking mechanism
that understands noise models of every device along the VC\@. Thus, the
``simpler'' protocol is implemented using an oracle: we use the simulation to
give us the fidelity. The QNP does not use this backdoor mechanism as it is not
available outside of simulations. We remark that \fref{fig:decoherence} shows
that the cutoff timer is more efficient than an end-node only strategy even
with the physically impossible direct access to the fidelity.

\begin{figure}[t!]
  \centering
  \includegraphics[width=0.9\linewidth]{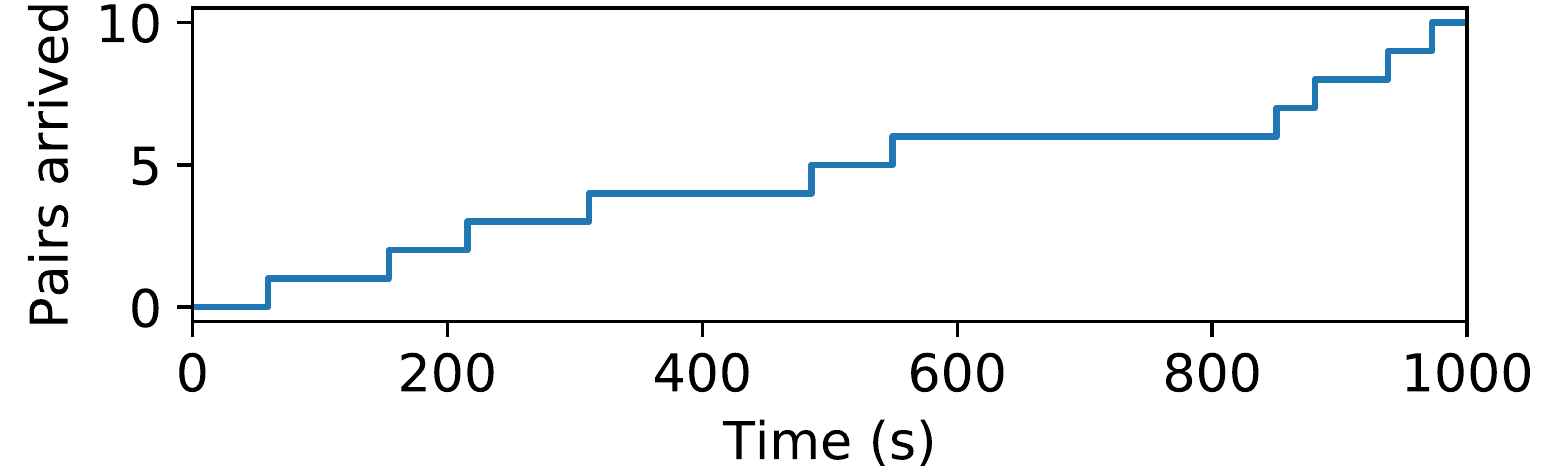}
  \caption{The number of pairs produced as a function of time on a near-future
    network. The protocol is able to deliver entanglement despite the limited
    resources.}\label{fig:demo}
\end{figure}

\textbf{Message delays} As memory lifetimes get shorter, the effect of message
delays becomes a concern. The QNP is designed such that quantum operations like
swapping never block waiting for control messages. To demonstrate the
effectiveness of this strategy in \fref{fig:decoherence}c we plot the
throughput of the two VCs as we introduce artificial delays to increase the
time between the sending of any QNP message to the moment that message is
processed at the next node. We perform the simulations for a memory lifetime of
about 1.6~s (approximately the middle of \fref{fig:decoherence}a) as it
corresponds to achievable lifetimes in current hardware~\cite{abobeih2018one}.
We see that the delay has no effect until it starts approaching the cutoff
timeout. Once classical control messages are delayed beyond this threshold the
delivered pairs have insufficient fidelity.

\subsection{Near-Future Hardware Performance}\label{sec:near-future}

So far, we considered a network that whilst not infeasible is still beyond our
capabilities. We demonstrate that the protocol remains functional even with
near-future hardware~\cite{abobeih2018one, humphreys2018deterministic} which
highlights the timeliness of our work (hardware model and parameters are
described in Appendix~\ref{app:hw}). \fref{fig:demo} shows the arrival times of
10 pairs requested over a linear network of three nodes with an inter-node
separation of 25\,km in a single simulation run. We request a fidelity of 0.5
which is sufficient to demonstrate quantum entanglement. In addition to more
realistic parameters there are other constraints. The nodes have only one
communication qubit and thus may only use one link at a time. As a result, a
pair must be moved into storage before another pair can be created to swap
with. Furthermore, the act of generating the next entangled pair further
degrades the stored qubits due to the dephasing of nuclear
spins~\cite{kalb2018dephasing}. Yet despite the enormous differences in the
operating environment the QNP remains functional: it exposes the right knobs to
the control plane which an operator that understands the limitations can
properly tune. As our routing protocol does not work well in this environment
we manually populate the routing tables. We set the link-fidelities as high as
possible to compensate for poor hardware quality and the nuclear dephasing and
we tune the cutoff timer to ensure we meet the end-to-end fidelity threshold.

\section{Discussion}

In this paper we have proposed a connection-oriented quantum data plane
protocol for delivering end-to-end entanglement across a quantum network.
However, whilst our work marks an important step on the way to large-scale
quantum networks it is only one component of a complete quantum network
architecture. Here, we briefly outline possible future directions of work.

\textbf{QNP services} We have designed the QNP using a VC approach inspired by
MPLS as a building block for more complex quantum network services such as the
entanglement distillation example described in Sec.~\ref{sec:distil}. Other
potential services include (i) services inspired by classical MPLS such as
multipath support or failure recovery and (ii) services that take advantage of
new features that are not present in classical networks such as the ability to
pre-generate and store entangled pairs in preparation for future
demand~\cite{chakraborty2019distributed}.

\textbf{Control plane design} In our paper we focused entirely on the quantum
data plane and considered only a simplified control plane. Control plane
protocols are also an emerging field in quantum network research, especially in
the area of routing~\cite{chakraborty2019distributed,
  chakraborty2020polynomial, caleffi2017optimal, gyongyosi2018decentralized,
  van2013path, schoute2016shortcuts, imre2012advanced,
  gyongyosi2017entanglement, li2020effective, shi2020concurrent}. However, more
work is needed for a complete quantum network control plane. In particular,
there is scope for further work on resource reservation, signalling, and more
generally traffic engineering in quantum networks. Furthermore, there is also
the question of software architecture for control planes: whether it is
distributed or centralised. For example, software-defined architectures have
been considered for QKD networks~\cite{aguado2020enabling} and more recently
have also been proposed for quantum repeater networks~\cite{kozlowski2020p4}.

\textbf{Relation to Internet protocol design} It has been shown that classical
network protocol stacks may be holistically analysed and systematically
designed as distributed solutions to some optimisation problems
(i.e.~generalised network utility maximisation)~\cite{chiang2007layering}. It
is conceivable that it is also possible to apply a similar ``Layering as
Optimisation Decomposition'' approach to the quantum network protocol stack to
improve its design.

\textbf{Heterogeneous networks} In this paper we focused on homogeneous
networks based on a single hardware platform as that is the focus for
near-future experimental work. However, a future quantum internet will
inevitably consist a wide variety of physical platforms resulting in very
different parameters for decoherence and quantum state fidelity for the quantum
nodes and links. Therefore, more work is needed to understand the performance
of quantum network protocols on hybrid quantum networks.

\section{Related Work}

\textbf{Quantum data plane protocols} Three other proposals for end-to-end
entanglement generation protocols that operate within our definition of a
quantum data plane have been put forward~\cite{yu2019protocols,
  huberman2019quantum, matsuo2019quantum}. Ref.~\cite{yu2019protocols} proposes
a scheme inspired by classical UDP/TCP based on quantum error correction which
is currently beyond hardware capabilities both in terms of required state
quality and number of qubits. Ref.~\cite{huberman2019quantum} does not consider
decoherence. Ref.~\cite{matsuo2019quantum} combines what we would define as a
quantum data plane protocol and a signalling protocol into one ``RuleSet''
based protocol, but the authors only study two-node networks with a single
link.

\textbf{Repeater chain protocols} Since many long-distance links in the quantum
internet will be built by chaining many quantum repeaters, protocols for such
constructions have received significant attention~\cite{van2008system,
  goebel2008multistage, briegel1998quantum, cirac1997quantum, dur1999quantum,
  duan2001long, santra2019quantum, brand2020efficient, khatri2019practical,
  rozpedek2018parameter, rozpedek2018near, sangouard2011quantum,
  munro2015inside}. However, these protocols are limited in scope to individual
chains: they cannot handle non-linear topologies and do not have mechanisms for
merging and splitting flows. Nevertheless, since a circuit in our network
protocol is in some ways like a repeater chain, we use many ideas from this
line of research, such as cutoff times~\cite{khatri2019practical,
  rozpedek2018parameter, rozpedek2018near, li2020efficient}.

\textbf{Network stacks} Our paper fits into the network stack architecture
proposed in Ref.~\cite{dahlberg2019link}. The authors in
Ref.~\cite{dahlberg2019link} have also designed a link layer protocol, but they
did not develop a network layer protocol. A complementary functional allocation
for a quantum network stack for entanglement distillation also
exists~\cite{aparicio2011protocol, van2012quantum, van2013designing} though no
concrete protocols have been given. An alternative outline for a quantum
network stack has also been put forward in Ref.~\cite{pirker2019quantum}, but
it does not account for many crucial low-level details such as hardware
imperfections or classical control.

\section{Conclusions}

In this paper we have taken another step towards large-scale quantum networks.
We have designed a quantum data plane network protocol for creating
long-distance end-to-end entangled pairs, the key resource for distributed
quantum applications. Quantum networks are complex systems and will require
sophisticated resource management and scheduling strategies. We designed the
Quantum Network Protocol to be the building block for constructing such
higher-level services much like MPLS and IP datagrams have been for classical
networks. We have ensured the protocol is efficient despite the extreme noise
intrinsic to quantum systems by leveraging virtual circuits, building upon a
robust link layer protocol, and adopting a cutoff timer. We also ensure that
our protocol is scalable and can remain usable in the future once more capable
hardware becomes available by leaving out tasks that require detailed knowledge
of the hardware parameters of the nodes and links in the network to supporting
protocols. This allows the core building block, the Quantum Network Protocol,
to remain the same whilst giving the control plane the flexibility to evolve
together with the network capabilities and requirements.

\begin{acks}

We would like to thank Matthew Skrzypczyk, Carlo Delle Donne, Przemys{\l}aw
Pawe{\l}czak, Tim Coopmans, and Bruno Rijsman for the many useful discussions
that helped us in this work and their detailed feedback on earlier versions of
this draft. We would also like to acknowledge Kaushik Chakraborty and Kenneth
Goodenough for further technical discussions.

The authors acknowledge funding received from the \grantsponsor{qt.flagship}{EU
  Flagship on Quantum Technologies}{https://qt.eu/}, \grantsponsor{qia}{Quantum
  Internet Alliance}{https://quantum-internet.team/}
(No.~\grantnum{qia}{820445}), an \grantsponsor{ercsg}{ERC Starting Grant}{}
(SW), and an \grantsponsor{nwovidi}{NWO VIDI Grant}{} (SW).

\end{acks}

\bibliographystyle{ACM-Reference-Format}
\bibliography{reference}

\clearpage
\appendix
\onecolumn

\section*{Appendix}

\section{Artifacts}

The source code for the implementation of the Quantum Network Protocol in
NetSquid~\cite{netsquid} and the raw data used to produce the plots in this
paper has been made available at
\url{https://doi.org/10.34894/2P1P91}~\cite{2P1P91_2020}.

\bigskip

The artifact directory contains:

\begin{itemize}
  \item the source code to run the simulations and reproduce the results,
  \item the raw data used to plot the figures in the paper.
\end{itemize}

\bigskip

The zip file contains a directory within which are contained:

\begin{itemize}
  \item README.md --- description of the contents as well as instructions to
    install and set up the simulations,
  \item EXPERIMENTS.md --- instructions to run the experiments described in
    this paper and reproduce all the data.
\end{itemize}

\bigskip

Note that compatibility on all platforms is not guaranteed. For this reason a
Dockerfile is also provided which should make it possible to execute the
artifacts on all platforms that support Docker containers. Instructions for
using the container are included in README.md.

\section{Hardware parameters}\label{app:hw}

The simulations in this paper are based on the nitrogen vacancy centre
(NV-centre) platform for quantum repeaters. Experimental results for this
platform are available in Refs.~\cite{abobeih2018one, bradley201910,
  kalb2017entanglement, taminiau2014universal, reiserer2016robust,
  cramer2016repeated, humphreys2018deterministic, riedel2017deterministic,
  zaske2012visible}. An in-depth introduction to the quantum physics and
operation of this platform including noise modelling and the definitions of the
different hardware parameters can be found in Appendix D of
Ref.~\cite{dahlberg2019link}. Parameter values used for simulations in this
paper are given in Tables~\ref{tab:gates} and~\ref{tab:params}. The near-term
values are based on references to the aforementioned experimental papers and
Ref.~\cite{dahlberg2019link}.

\textbf{Simulation parameters} All of the simulations in the paper except for
the near-future hardware example were done in an optimistic configuration with
hardware parameters beyond what is currently possible in the laboratory. These
parameters are shown in Tables~\ref{tab:gates} and~\ref{tab:params} where they
are also compared to the currently achievable parameters. Additionally, we made
a few simplifications that go beyond hardware parameter values.

We did not distinguish between so-called communication (electron) qubits and
memory (carbon) qubits. In an NV-centre architecture only one qubit, the
communication qubit, can participate in link-pair generation at any one time.
This means that only one link of every node can be active at any one time. The
quantum network protocol, as demonstrated in the near-future hardware
simulations, can cope with this scenario, but for larger networks requires a
more sophisticated resource management and scheduling approach which is beyond
the scope of this work. Therefore, for the purposes of our simulations (except
for the near-future hardware case) all qubits are treated as communication
(electron) qubits and can participate in link-pair generation.

Furthermore, a major source of noise in NV-centres is the dephasing of nuclear
spins (memory qubits) due to the resetting of the communication qubit during
entanglement generation attempts. Since we only consider communication qubits
in our simulations we also do not consider this noise in our simulations.
However, from the point of view of our protocol this noise can be treated like
normal decoherence --- it is a process that degrades the quality of idle qubits
over time. Nevertheless, this requires a more sophisticated approach to
correctly calculate the cutoff timeout values for idle qubits which is also
beyond the scope of this paper. However, our near-future hardware example in
the main text, where we hand-picked a timeout value, shows that the cutoff time
of the protocol is a suitable mechanism for handling this noise.

\textbf{Optical fibres} The channels that carry photons and classical messages
between the nodes (both classical and quantum channels) are standard telecom
optical fibres. For the near-term hardware simulation we considered fibres of
25~km length between each node which requires frequency conversion for the
photons used in entanglement generation (to achieve $0.5$~db/km losses). For
the rest of the simulations we used parameters closer to a lab scenario, 2~m
fibres, as they do not need frequency conversion (losses of $5$~dB/km) leading
to faster generation rates. We do not simulate losses for classical messages,
because (i) they are extremely low, (ii) protocol communication happens over
TCP so lost packets would just be resent, (iii) in the main text we already
consider the effects of arbitrary processing and communication delays which can
arise from TCP retransmission.

\begin{table*}[p]
  \begin{center}
    \begin{tabular}{ l c c c c }
      & \multicolumn{2}{c}{Simulation} & \multicolumn{2}{c}{Near-term (\fref{fig:demo})} \\
      & Fidelity & Duration & Fidelity & Duration \\
      \hline
      Electron single-qubit gate                                       & 1.0   & 5~ns       & 1.0   & 5~ns       \\
      Two-qubit gate (E-C controlled $\sqrt{\chi}$-gate for near-term) & 0.998 & 500~$\mu$s & 0.992 & 500~$\mu$s \\
      Carbon Rot-Z gate                                                & ---   & ---        & 1.0   & 20~$\mu$s  \\
      Electron initialisation in $\ket{0}$                             & 0.99  & 2~$\mu$s   & 0.99  & 2~$\mu$s   \\
      Carbon initialisation in $\ket{0}$                               & ---   & ---        & 0.95  & 300~$\mu$s \\
      Electron readout $\ket{0}$                                       & 0.998 & 3.7~$\mu$s & 0.95  & 3.7~$\mu$s \\
      Electron readout $\ket{1}$                                       & 0.998 & 3.7~$\mu$s & 0.995 & 3.7~$\mu$s \\
      &&&& \\
    \end{tabular}
    \caption{Quantum gate parameters. Explanation of each parameter can be
      found in Appendix D of Ref.~\cite{dahlberg2019link}.}\label{tab:gates}
  \end{center}
\end{table*}

\begin{table*}[p]
  \begin{center}
    \begin{tabular}{ l c c }
      & Simulation & Near-term (\fref{fig:demo}) \\
      \hline
      Electron $T_1$                  & $>$1~h               & $>$1~h                \\
      Electron $T^*_2$                & 60~s                 & 1.46~s                \\
      Carbon $T_1$                    & ---                  & $>$ 6~m               \\
      Carbon $T^*_2$                  & ---                  & 60~s                  \\
      $\Delta \omega$                 & ---                  & $2\pi \times 377$~kHz \\
      $\tau_d$                        & ---                  & 82~ns                 \\
      $\tau_w$                        & 25~ns                & 25~ns                 \\
      $\tau_e$                        & 6.0~ns               & 6.48~ns               \\
      $\Delta \phi$                   & 2.0$^\text{\degree}$ & 10.6$^\text{\degree}$ \\
      $p_\text{double\_excitation}$   & 0.00                 & 0.04                  \\
      $p_\text{zero\_phonon}$         & 0.75                 & 0.46                  \\
      Collection efficiency           & $20.0 \cdot 10^{-3}$ & $4.38 \cdot 10^{-3}$  \\
      Dark count rate                 & 20 s$^{-1}$          & 20 s$^{-1}$           \\
      $p_\text{detection}$            & 0.8                  & 0.8                   \\
      Visibility (distinguishability) & 1.0                  & 0.9                  \\
      && \\
    \end{tabular}
    \caption{Other hardware parameters. Explanation of each parameter can be
      found in Appendix D of Ref.~\cite{dahlberg2019link}.}\label{tab:params}
  \end{center}
\end{table*}

\clearpage

\section{Protocol description --- NOT INCLUDED IN CoNEXT SUBMISSION}

\subsection{Identifiers}

\textbf{Circuit ID} As circuits are the responsibility of the signalling
protocol, the quantum network protocol treats the circuit ID as an opaque
handle which it includes in its messages to identify which circuit they pertain
to.

\textbf{Address} An address uniquely identifies a communication end-point in
the quantum network. We use the locator/identifier scheme for addressing and
thus an address consists of a locator and an identifier. A \textbf{locator} is
a network-wide unique handle that identifies a quantum node for the purposes of
routing. An \textbf{identifier} specifies a unique communication end-point on a
particular node. We do not specify the exact format of these values. For
example, one could use IP addresses for the locator and port numbers for the
identifier. In our simulations, we used strings for locators and integers for
identifiers.

\textbf{Link-pair correlator} The link-pair correlator uniquely identifies a
pair generated on a particular link. The network protocol uses this correlator
to identify the qubits it wishes to operate on to the local node. Therefore,
the requirements on this correlator are: (i) it is delivered by the link layer
protocol together with the pair and (ii) the pair of nodes that generated this
pair must be able to map the correlator to the appropriate qubits in their
local memory. It is not required that this correlator be meaningful beyond a
pair of nodes that share the link. Our simulations use the link layer
protocol's entanglement identifier for this purpose. This identifier is a
three-tuple of \texttt{(node-id-1, node-id-2, sequence-number)}.

\textbf{Link-label} A single link-label is allocated for each link on the
circuit's path by the signalling protocol. It is used to identify requests to
the link layer dedicated to that particular circuit and conceptually are
similar to MPLS labels. It should have a unique 1-to-1 mapping to a circuit ID,
but this mapping can be different on each link along the path. It is required
that the link layer protocol be able to deliver the link-pairs together with
their circuit label at the two ends of the link. Our simulations use the link
layer protocol's purpose identifier to achieve this.

\textbf{Request ID} A request ID uniquely identifies a request between a pair
of addresses. It is assigned by the application using the network protocol. In
case of a duplicate request ID being issued, the protocol will reject it. The
purpose of the request ID is to allow application to reuse an address end-point
(locator+identifier) for multiple requests.

\subsection{Messages}\label{app:msg}

The quantum network protocol has two groups of messages operating at two levels
of granularity: per-request, and per-pair.

\vspace{0.5em} \textbf{Request level} These messages operate at an individual
request level within a particular circuit. A request is between two
communication end-points for some number or rate of entangled pairs. Multiple
requests can share a circuit.

\begin{itemize}
  \item \textbf{FORWARD} --- The FORWARD message propagates a request from the
    head-end of the circuit to the tail-end. The information it carries is used
    for two purposes: (i) to initiate/update link layer requests and (ii) to
    provide the tail-end with enough information for its book-keeping. It has
    the following structure:
    \vspace{-0.5em}
    \begin{verbatim}
FORWARD:
  circuit_id
  request_id
  head_end_identifier
  tail_end_identifier
  request_type
  number_of_pairs
  final_state
  rate
    \end{verbatim}
    \vspace{-1.5em}
    where the different values are defined as:
    \begin{itemize}[noitemsep]
      \item \texttt{circuit\_id} --- the opaque circuit ID,
      \item \texttt{request\_id} --- the individual request ID,
      \item \texttt{head\_end\_identifier} --- the end-point address identifier
        at the head-end node,
      \item \texttt{tail\_end\_identifier} --- the end-point address identifier
        at the tail-end node.
      \item \texttt{request\_type} --- NORMAL/EARLY/MEASURE, indicates when the
        pair is to be consumed.
      \item \texttt{number\_of\_pairs} --- the number of pairs in this request
        (left unspecified for rate requests),
      \item \texttt{final\_state} --- set if the request specified a particular
        Bell state it wishes its pairs to be delivered in.
      \item \texttt{rate} --- the end-to-end rate (EER) that the sum of all
        active requests require from the circuit.
    \end{itemize}
    The \texttt{circuit\_id} is used for associating the message with the right
    circuit. The next four values are used by the tail-end node for request
    book-keeping. The \texttt{request\_type} indicates when the pair is to be
    consumed and it takes one of three values: NORMAL/EARLY/MEASURE\@. NORMAL
    pairs are to be delivered once successful creation is confirmed with a
    tracking message. EARLY pairs are delivered as soon as a qubit is available
    at the end-node, but the application must take over the responsibility of
    handling unsuccessful pairs. Tracking information is delivered once
    available for post-processing. MEASURE pairs have their qubits immediately
    measured by the QNP\@. The result is withheld until tracking information is
    available so that only outcomes from successfully generated pairs are
    delivered. The \texttt{final\_state} indicates if a Pauli correction is to
    be made. If a user request wants the pairs delivered in a particular Bell
    state this field indicates which state it is. If it is set, the head-end
    node will perform the Pauli correction once it receives the entanglement
    tracking information. Finally, the \texttt{rate} is calculated by the
    head-end node based on all active requests it is serving. It is used to
    calculate the new LPR for link layer protocol. The network layer requests a
    fraction of the link's maximum LPR equal to the fraction of the circuit's
    EER that it needs. The routing table entry will contain the link's max LPR
    and the circuit's max EER for this calculation. However, note that it is
    the responsibility of the resource reservation, policing, and shaping
    mechanisms to ensure that this does not overload the link and that this
    rate is achievable.

  \item \textbf{COMPLETE} The COMPLETE message is the reverse of FORWARD and
    also propagates from the head-end to the tail-end of the circuit. Its
    purpose is to: (i) update/terminate link layer requests and (ii) notify the
    tail-end of a request's completion. It has the following structure:
    \vspace{-0.5em}
    \begin{verbatim}
COMPLETE:
  circuit_id
  request_id
  head_end_identifier
  tail_end_identifier
  rate
    \end{verbatim}
    \vspace{-1.5em}
    where the different values are defined in the same way as for FORWARD\@.
\end{itemize}

\vspace{0.5em} \textbf{Pair level} The next level operates on a per-pair
granularity as it tracks individual entangled pairs across the circuit.

\begin{itemize}
  \item \textbf{TRACK} --- The TRACK message is the key quantum data plane
    message of the entire protocol. It tracks the chain of link-pairs and their
    connecting entanglement swaps across the entire circuit whilst also
    collecting information necessary to infer the final state of the end-to-end
    pair. The TRACK message is sent in both directions of the circuit to avoid
    the need to acknowledge its receipt. It has the following structure:
    \vspace{-0.5em}
    \begin{verbatim}
TRACK:
  circuit_id
  request_id
  head_end_identifier
  tail_end_identifier
  origin_correlator
  link_correlator
  outcome_state
  epoch
    \end{verbatim}
    \vspace{-1.5em}
    where the different values are defined as:
    \begin{itemize}[noitemsep]
      \item \texttt{circuit\_id} --- the opaque circuit ID,
      \item \texttt{request\_id} --- the individual request ID,
      \item \texttt{head\_end\_identifier} --- the end-point address identifier
        at the head-end node,
      \item \texttt{tail\_end\_identifier} --- the end-point address identifier
        at the tail-end node,
      \item \texttt{origin\_correlator} --- the correlator for the link-pair
        that begins the chain,
      \item \texttt{link\_correlator} --- the correlator for the link-pair that
        continues the chain (updated at each node),
      \item \texttt{outcome\_state} --- the estimated state of the end-to-end
        based on information collected so far (updated at each node),
      \item \texttt{epoch} --- set by the head-end node the epoch to use after
        delivering this pair.
    \end{itemize}
    The \texttt{circuit\_id} identifies the circuit to which the tracked pair
    belongs to. The next three values uniquely identify the communication
    end-points that will receive the pair. These values are set by the
    demultiplexing algorithm. The \texttt{origin\_correlator} specifies the
    link-pair, and thus the qubit, that belongs to this end-to-end pair at the
    origin node of the message. It is used for the EXPIRE message. The
    \texttt{link\_correlator} is updated at every node and at the final node of
    the circuit will identify the qubit that belongs to the pair specified by
    the \texttt{request\_id} and \texttt{sequence}. The \texttt{outcome\_state}
    is also updated at every node and at the end of the circuit will identify
    which of the four Bell states the final end-to-end pair is in. The
    \texttt{epoch} is set by the head-end and indicates the epoch (set of
    active requests) that activates after this pair is delivered.

  \item \textbf{EXPIRE} --- The EXPIRE message serves to notify the end-nodes
    that the chain they originate was broken by an expired qubit. Only the
    intermediate nodes are free to discard qubits when their expiry timer pops.
    They can do this, because after discarding a qubit they can simply wait for
    the TRACK message to turn it around. They don't have to worry about what
    happens if the timer pops after a swap has already happened, because at
    that point those qubits are no longer part of the end-to-end chain.
    However, the end-nodes do not have the same flexibility. The qubits at the
    end-nodes are not entanglement swapped and they are the qubits that are
    eventually delivered to the application. The end-nodes will have sent their
    TRACK message as soon as their link-pairs were generated. However, the two
    qubits of the final end-to-end pair are not delivered at the same --- they
    are delivered on receipt of a TRACK message. This means that a timer-based
    expiry mechanism at the end-nodes may run into a window condition where one
    end has already been delivered to the application, but the other end is
    discarded due to a popped timer. Therefore, end-nodes are only allowed to
    discard their qubits on receipt of an EXPIRE message, because only then can
    they be sure that the other end will not be delivered. It has the following
    structure:
    \vspace{-0.5em}
    \begin{verbatim}
EXPIRE:
  circuit_id
  origin_correlator
    \end{verbatim}
    \vspace{-1.5em}
    where the different values are defined in the same way as for TRACK\@.
\end{itemize}

\subsection{Rules}

The protocol executes rules in response to received classical messages, such as
TRACK and EXPIRE, and link-pairs from the link layer. The actions taken in
response to FORWARD and COMPLETE are self-explanatory from the nature of the
messages, but TRACK, EXPIRE, and link-pair rules are more involved and thus we
explain them in detail in the following section. The rules are different
depending on whether the node is an end-node or an intermediate node which
simply follows from the fact that only intermediate nodes execute entanglement
swaps. However, head-end and tail-end rules also differ, because the head-end
node has some additional management responsibilities.

\vspace{0.5em} \textbf{Head-end rules} The head-end rules are responsible for
initiating and handling TRACK messages as well as handling EXPIRE messages. The
head-end node is also responsible for advancing epochs, performing Pauli
corrections if required, and sending COMPLETE messages for finished requests.

The head-end node has three rules:
\begin{enumerate}
  \item LINK rule: triggered whenever the head-end node receives a link-pair
    from the link layer protocol with a link label assigned to this circuit.
    Shown in Alg.~\ref{alg:head-link}.
  \item TRACK rule: triggered on every TRACK message received by the head-end
    node for this circuit. Shown in Alg.~\ref{alg:head-track}.
  \item EXPIRE rule: triggered on every EXPIRE message received by the head-end
    node for this circuit. Shown in Alg.~\ref{alg:head-expire}.
\end{enumerate}

\vspace{0.5em} \textbf{Tail-end rules} The tail-end rules are similar to the
head-end rules, but have fewer management responsibilities. Mainly, it does not
initiate FORWARD/COMPLETE messages, and it relies on the head-end to advance
the epochs through TRACK messages.

The tail-end node also has three rules:
\begin{enumerate}
  \item LINK rule: triggered whenever the tail-end node receives a link-pair
    from the link layer protocol with a link label assigned to this circuit.
    Shown in Alg.~\ref{alg:tail-link}.
  \item TRACK rule: triggered on every TRACK message received by the tail-end
    node for this circuit. Shown in Alg.~\ref{alg:tail-track}.
  \item EXPIRE rule: triggered on every EXPIRE message received by the tail-end
    node for this circuit. Shown in Alg.~\ref{alg:tail-expire}.
\end{enumerate}

\vspace{0.5em} \textbf{Entanglement swap rules} The entanglements swap rules
are conceptually the simplest. As soon as two pairs, one upstream and one
downstream, are available the node must execute an entanglement swap. There are
some additional considerations to handle qubit expiry and correctly handling of
TRACK messages depending on whether they arrive before or after the
entanglement swap.

\begin{enumerate}
  \item LINK rule: triggered whenever the intermediate node receives a
    link-pair from the link layer protocol with a link label assigned to this
    circuit. Shown in Alg.~\ref{alg:int-link}.
  \item TRACK rule: triggered on every TRACK message received by the
    intermediate node for this circuit. Shown in Alg.~\ref{alg:int-track}.
  \item Expire rule: triggered when a qubit's cutoff timer pops. Shown in
    Alg.~\ref{alg:int-expire}.
\end{enumerate}

\vspace{0.5em} \textbf{Demultiplexing} The protocol uses a demultiplexer at
both end-nodes that assigns VC pairs to requests. This can be done
symmetrically (both end-nodes consistently pick a request) or asymmetrically
(one end-node chooses and communicates to other end-node). In the case of
symmetric demultiplexing, we allow for occasional inconsistent decisions, but
the demultiplexer must then perform cross-checks to discard such pairs. This is
useful since qubits can be discarded within the network due to the cutoff timer
mechanism making it difficult for the end-nodes to know how many qubits have
been generated and discarded at the other end of the VC\@. Symmetric
demultiplexing is required for EARLY and MEASURE request types.

In our simulations we have implemented a symmetric demultiplexer that assigns
MEASURE request pairs as soon as the pair starts generating, EARLY and NORMAL
requests pairs as soon as the end-node's link-pair completes. However, the
demultiplexer can reassign NORMAL requests at the tail-end and use the
head-end's decisions when an inconsistent decision is spotted as the pair has
not been delivered yet. This is also possible for MEASURE requests if the
mis-matching head-end and tail-end requests are for the same measurement since
the outcomes are withheld until the tracking information is delivered. For
EARLY requests it is only possible to deliver an failure notification.

\vspace{0.5em} \textbf{Objects} In the algorithm listings below we will refer
to the following two components from \fref{fig:system}:

\begin{itemize}
  \item \texttt{demultiplexer} --- the demultiplexer.
  \item \texttt{qsched} --- quantum task scheduler.
  \item \texttt{qmm} --- quantum memory manager.
\end{itemize}

\textbf{Functions} In the algorithm listings below we refer several helper
unctions. Here we define ones whose role may be difficult to infer from the
name:

\begin{itemize}
  \item \texttt{pauli\_correction} --- transform the provided entangled pair
    from its current state into a desired Bell state.
  \item \texttt{entanglement\_swap} --- performs an entanglement swap by means
    of a Bell state measurement and returns a two-bit result of the
    measurement.
  \item \texttt{combine\_state} --- calculates the resulting Bell state after
    an entanglement swap given the input states and the two-bit swap outcome
    mapped to the Bell state it corresponds to.
\end{itemize}

\newpage
\begin{algorithm}[H]
  \SetAlgoLined{}
  \DontPrintSemicolon{}
  \SetKwInOut{Input}{input}
  \Input{link\_qubit --- qubit from entangled pair produced the link layer}
  \BlankLine{}
  request $\leftarrow$ demultiplexer.next\_request\@()\;
  \BlankLine{}
  TRACK $\leftarrow$ empty\_track\_message\@()\;
  TRACK.circuit\_id $\leftarrow$ circuit\_id\;
  TRACK.request\_id $\leftarrow$ request.id\;
  TRACK.head\_end\_identifier $\leftarrow$ request.head\_end.address.identifier\;
  TRACK.tail\_end\_identifier $\leftarrow$ request.tail\_end.address.identifier\;
  TRACK.origin\_correlator $\leftarrow$ link\_qubit.pair.correlator\;
  TRACK.link\_correlator $\leftarrow$ link\_qubit.pair.correlator\;
  TRACK.outcome\_state $\leftarrow$ link\_qubit.pair.state\;
  TRACK.epoch $\leftarrow$ get\_next\_epoch\@()\;
  \BlankLine{}
  Send TRACK to downstream node\;
  in\_transit[link\_qubit.pair.correlator] $\leftarrow$ request\;
  \caption{Head-end LINK rule}\label{alg:head-link}
\end{algorithm}

\begin{algorithm}[H]
  \SetAlgoLined{}
  \DontPrintSemicolon{}
  \SetKwInOut{Input}{input}
  \Input{TRACK --- a TRACK message received from downstream node}
  \BlankLine{}
  qubit $\leftarrow$ qmm.get\@(TRACK.link\_correlator)\;
  request $\leftarrow$ in\_transit.pop\@(TRACK.link\_correlator)\;
  state $\leftarrow$ TRACK.outcome\_state\;
  \BlankLine{}
  \If{demultiplexer.cross\_check\@(request, TRACK) fails} {
    qmm.free\@(TRACK.link\_correlator)\;
    return\;
  }
  \BlankLine{}
  \If{request.state is not null}{
    qsched.pauli\_correction\@(qubit, state, request.state)\;
    state $\leftarrow$ request.state\;
  }
  \BlankLine{}
  deliver (qubit, state, request.id, request.next\_sequence\@()) to request.head\_end.address.identifier
  \BlankLine{}
  \If{request is complete}{
    COMPLETE $\leftarrow$ empty\_complete\_message\@()\;
    COMPLETE.circuit\_id $\leftarrow$ circuit\_id\;
    COMPLETE.request\_id $\leftarrow$ request.id\;
    COMPLETE.head\_end\_identifier $\leftarrow$ request.head\_end.address.identifier\;
    COMPLETE.tail\_end\_identifier $\leftarrow$ request.tail\_end.address.identifier\;
    COMPLETE.link\_mean\_rate $\leftarrow$ calculate\_new\_eer\@()\;
    \BlankLine{}
    Send COMPLETE to downstream node\;
  }
  \caption{Head-end TRACK rule}\label{alg:head-track}
\end{algorithm}

\begin{algorithm}[H]
  \SetAlgoLined{}
  \DontPrintSemicolon{}
  \SetKwInOut{Input}{input}
  \Input{EXPIRE --- an EXPIRE message received from downstream node}
  \BlankLine{}
  qmm.free\@(EXPIRE.origin\_correlator)\;
  Clear in\_transit[EXPIRE.origin\_correlator]\;
  \caption{Head-end EXPIRE rule}\label{alg:head-expire}
\end{algorithm}

\newpage
\begin{algorithm}[H]
  \SetAlgoLined{}
  \DontPrintSemicolon{}
  \SetKwInOut{Input}{input}
  \Input{link\_qubit --- qubit from entangled pair produced the link layer}
  \BlankLine{}
  request $\leftarrow$ demultiplexer.next\_request\@()\;
  \BlankLine{}
  TRACK $\leftarrow$ empty\_track\_message\@()\;
  TRACK.circuit\_id $\leftarrow$ circuit\_id\;
  TRACK.request\_id $\leftarrow$ request.id\;
  TRACK.head\_end\_identifier $\leftarrow$ request.head\_end.address.identifier\;
  TRACK.tail\_end\_identifier $\leftarrow$ request.tail\_end.address.identifier\;
  TRACK.origin\_correlator $\leftarrow$ link\_qubit.pair.correlator\;
  TRACK.link\_correlator $\leftarrow$ link\_qubit.pair.correlator\;
  TRACK.outcome\_state $\leftarrow$ link\_qubit.pair.state\;
  TRACK.epoch $\leftarrow$ null\;
  \BlankLine{}
  Send TRACK to upstream node\;
  in\_transit[link\_qubit.pair.correlator] $\leftarrow$ request\;
  \caption{Tail-end LINK rule}\label{alg:tail-link}
\end{algorithm}

\begin{algorithm}[H]
  \SetAlgoLined{}
  \DontPrintSemicolon{}
  \SetKwInOut{Input}{input}
  \Input{TRACK --- a TRACK message received from downstream node}
  \BlankLine{}
  qubit $\leftarrow$ qmm.get\@(TRACK.link\_correlator)\;
  request $\leftarrow$ in\_transit.pop\@(TRACK.link\_correlator)\;
  state $\leftarrow$ TRACK.outcome\_state\;
  \BlankLine{}
  \If{demultiplexer.cross\_check\@(request, TRACK) fails} {
    qmm.free\@(TRACK.link\_correlator)\;
    return\;
  }
  \BlankLine{}
  \If{request.state is not null}{
    state $\leftarrow$ request.state\;
  }
  \BlankLine{}
  deliver (qubit, state, request.id, request.next\_sequence\@()) to request.tail\_end.address.identifier
  \caption{Tail-end TRACK rule}\label{alg:tail-track}
\end{algorithm}

\begin{algorithm}[H]
  \SetAlgoLined{}
  \DontPrintSemicolon{}
  \SetKwInOut{Input}{input}
  \Input{EXPIRE --- an EXPIRE message received from upstream node}
  \BlankLine{}
  qmm.free\@(EXPIRE.origin\_correlator)\;
  \caption{Tail-end EXPIRE rule}\label{alg:tail-expire}
\end{algorithm}

\newpage
\begin{algorithm}[H]
  \SetAlgoLined{}
  \DontPrintSemicolon{}
  upstream\_qubit $\leftarrow$ get\_next\_upstream\_qubit\@()\;
  downstream\_qubit $\leftarrow$ get\_next\_downstream\_qubit\@()\;
  \BlankLine{}
  \If{upstream\_qubit and downstream\_qubit are not null}{
    swap\_outcome\_state $\leftarrow$ qsched.entanglement\_swap\@(upstream\_qubit, downstream\_qubit)\;
    \BlankLine{}
    \eIf{upstream\_track[upstream\_qubit.pair.correlator] is not null}{
      TRACK $\leftarrow$ upstream\_track.pop\@(upstream\_qubit.pair.correlator)\;
      \BlankLine{}
      TRACK.link\_correlator $\leftarrow$ downstream\_qubit.pair.correlator\;
      TRACK.outcome\_state $\leftarrow$
      combine\_state\@(TRACK.outcome\_state, downstream\_qubit.pair.state, swap\_outcome\_state)\;
      \BlankLine{}
      Forward TRACK message to downstream node\;
      Clear all upstream\_expire\_record contents\;
    }{
      upstream\_qubit\_record[upstream\_qubit.pair.correlator] $\leftarrow$
      (downstream\_qubit, swap\_outcome\_state)\;
    }
    \BlankLine{}
    \eIf{downstream\_track[downstream\_qubit.pair.correlator] is not null}{
      TRACK $\leftarrow$ downstream\_track.pop\@(downstream\_qubit.pair.correlator)\;
      \BlankLine{}
      TRACK.link\_correlator $\leftarrow$ upstream\_qubit.pair.correlator\;
      TRACK.outcome\_state $\leftarrow$
      combine\_state\@(TRACK.outcome\_state, upstream\_qubit.pair.state, swap\_outcome\_state)\;
      \BlankLine{}
      Forward TRACK message to upstream node\;
      Clear all downstream\_expire\_record contents\;
    }{
      downstream\_qubit\_record[downstream\_qubit.pair.correlator] $\leftarrow$
      (upstream\_qubit, swap\_outcome\_state)\;
    }
  }
  \caption{Intermediate node LINK rule}\label{alg:int-link}
\end{algorithm}

\newpage
\begin{algorithm}[H]
  \SetAlgoLined{}
  \DontPrintSemicolon{}
  \SetKwInOut{Input}{input}
  \Input{TRACK --- a TRACK message received from the downstream or upstream
    node}
  \BlankLine{}
  \uIf{TRACK received from upstream node}{
    \uIf{upstream\_qubit\_record[TRACK.link\_correlator] is not null}{
      (downstream\_qubit, swap\_outcome\_state) $\leftarrow$
      upstream\_qubit\_record.pop\@(TRACK.link\_.correlator)\;
      \BlankLine{}
      TRACK.link\_correlator $\leftarrow$ downstream\_qubit.pair.correlator\;
      TRACK.outcome\_state $\leftarrow$
      combine\_state\@(TRACK.outcome\_state, downstream\_qubit.pair.state, swap\_outcome\_state)\;
      \BlankLine{}
      Forward TRACK message to downstream node\;
    }\uElseIf{upstream\_expire\_record[TRACK.link\_correlator] is not null}{
      EXPIRE $\leftarrow$ empty\_expire\_message\@()\;
      EXPIRE.circuit\_id $\leftarrow$ TRACK.circuit\_id\;
      EXPIRE.origin\_correlator $\leftarrow$ TRACK.origin\_correlator
      \BlankLine{}
      Send EXPIRE message upstream to head-end node.\;
      Clear upstream\_expire\_record[TRACK.link\_correlator]\;
    }\Else{
      upstream\_track[TRACK.link\_correlator] $\leftarrow$ TRACK\;
    }
  }
  \ElseIf{TRACK received from downstream node}{
    \uIf{downstream\_qubit\_record[TRACK.link\_correlator] is not null}{
      (upstream\_qubit, swap\_outcome\_state) $\leftarrow$
      downstream\_qubit\_record.pop\@(TRACK.link\_correlator)\;
      \BlankLine{}
      TRACK.link\_correlator $\leftarrow$ upstream\_qubit.pair.correlator\;
      TRACK.outcome\_state $\leftarrow$
      combine\_state\@(TRACK.outcome\_state, upstream\_qubit.pair.state, swap\_outcome\_state)\;
      \BlankLine{}
      Forward TRACK message to upstream node\;
    }\uElseIf{downstream\_expire\_record[TRACK.link\_correlator] is not null}{
      EXPIRE $\leftarrow$ empty\_expire\_message\@()\;
      EXPIRE.circuit\_id $\leftarrow$ TRACK.circuit\_id\;
      EXPIRE.origin\_correlator $\leftarrow$ TRACK.origin\_correlator
      \BlankLine{}
      Send EXPIRE message upstream to head-end node.\;
      Clear downstream\_expire\_record[TRACK.link\_correlator]\;
    }\Else{
      downstream\_track[TRACK.link\_correlator] $\leftarrow$ TRACK\;
    }
  }
  \caption{Intermediate node TRACK rule}\label{alg:int-track}
\end{algorithm}

\newpage
\begin{algorithm}[H]
  \SetAlgoLined{}
  \DontPrintSemicolon{}
  \SetKwInOut{Input}{input}
  \Input{expired\_qubit --- qubit whose expiry timeout just popped}
  \BlankLine{}
  \uIf{expired\_qubit is from upstream link}{
    \eIf{upstream\_track[expired\_qubit.pair.correlator] is not null}{
      TRACK $\leftarrow$ upstream\_track.pop\@(expired\_qubit.pair.correlator)\;
      \BlankLine{}
      EXPIRE $\leftarrow$ empty\_expire\_message\@()\;
      EXPIRE.circuit\_id $\leftarrow$ TRACK.circuit\_id\;
      EXPIRE.origin\_correlator $\leftarrow$ TRACK.origin\_correlator
      \BlankLine{}
      Send EXPIRE message upstream to head-end node\;
    }{
      upstream\_expire\_record[expired\_qubit.pair.correlator] $\leftarrow$ expired\_qubit\;
    }
  }\ElseIf{expired\_qubit is from downstream link}{
    \eIf{downstream\_track[expired\_qubit.pair.correlator] is not null}{
      TRACK $\leftarrow$ downstream\_track.pop\@(expired\_qubit.pair.correlator)\;
      \BlankLine{}
      EXPIRE $\leftarrow$ empty\_expire\_message\@()\;
      EXPIRE.circuit\_id $\leftarrow$ TRACK.circuit\_id\;
      EXPIRE.origin\_correlator $\leftarrow$ TRACK.origin\_correlator
      \BlankLine{}
      Send EXPIRE message downstream to tail-end node\;
    }{
      downstream\_expire\_record[expired\_qubit.pair.correlator] $\leftarrow$ expired\_qubit\;
    }
  }
  \caption{Intermediate node expiry rule}\label{alg:int-expire}
\end{algorithm}

\end{document}